\documentclass[twocolumn,twocolappendix]{aastex7}

\usepackage{amsmath}
\usepackage{amssymb}
\usepackage{CJK}
\usepackage{booktabs}
\usepackage{graphicx}
\usepackage{caption}
\usepackage{subcaption}
\usepackage{graphicx}

\newcommand\egg{\textit{the Egg}}

\begin{document}

\title{Synthetic Spectral Library of Optically Thick Atmospheres for Little Red Dots}

\begin{CJK*}{UTF8}{gbsn}

\author[0000-0003-2488-4667]{Hanpu Liu (刘翰溥)}
\affiliation{Department of Astrophysical Sciences, Princeton University, 4 Ivy Lane, Princeton, NJ 08544, USA}
\email[show]{hanpu.liu@princeton.edu}

\author[0000-0002-2624-3399]{Yan-Fei Jiang (姜燕飞)}
\affiliation{Center for Computational Astrophysics, Flatiron Institute, New York, NY 10010, USA}
\email{yjiang@flatironinstitute.org}

\author[0000-0001-9185-5044]{Eliot Quataert}
\affiliation{Department of Astrophysical Sciences, Princeton University, 4 Ivy Lane, Princeton, NJ 08544, USA}
\email{quataert@princeton.edu}

\author[0000-0002-5612-3427]{Jenny E. Greene}
\affiliation{Department of Astrophysical Sciences, Princeton University, 4 Ivy Lane, Princeton, NJ 08544, USA}
\email{jgreene@astro.princeton.edu}

\author[0000-0002-0463-9528]{Yilun Ma (马逸伦)}
\affiliation{Department of Astrophysical Sciences, Princeton University, 4 Ivy Lane, Princeton, NJ 08544, USA}
\email{yilun@princeton.edu}

\author[0000-0001-6052-4234]{Xiaojing Lin}
\affiliation{Department of Astronomy, Tsinghua University, Beijing 100084, China}
\email{xiaojinglin.astro@gmail.com}

\begin{abstract}
Little Red Dots (LRDs) challenge conventional models of active galactic nuclei. At rest-optical-to-near-infrared (IR) wavelengths, these compact extragalactic objects show blackbody-like continuum emission and spectral features reminiscent of stars, motivating models with an optically thick atmosphere at $T_{\rm\!\,eff}\sim4000-5000{\rm~K}$. We develop (and publicly release) a synthetic spectral library of optically thick atmospheres with gas conditions tailored for LRDs, parameterized by effective temperature $T_{\rm\!\,eff}$ and surface gravity $g$. Given the uncertain dynamical structure of LRDs, we interpret $g$ mainly as a proxy for the photospheric density $\rho_{\rm\!\,ph}$. We show that blackbodies are only crude approximations to the emission from LRD-like atmospheres. Spectral features are abundant, many of which are sensitive diagnostics of photospheric density, including the overall curvature of the continuum, the rest-$1.6{\rm~\mu m}$ ``kink'' from $\rm H^-$ opacity, and the Ca~II triplet (CaT) absorption at rest-$8500{\rm~\AA}$. When compared against a local LRD, \egg, all three features are consistent with a low photospheric density $\rho_{\rm ph}\sim10^{-11}{\rm~g~cm^{-3}}$ ($g\sim10^{-3}{\rm~cm~s^{-2}}$ in our library), although CaT alone admits another higher-density solution. This low $\rho_{\rm ph}$ directly results from our radiative transfer modeling; with the additional assumption that the CaT line width traces turbulent support at the continuum photosphere in a spherical geometry, we infer a mass within the photosphere (black hole plus gas) of $\sim10^4~M_\odot$, with an Eddington ratio $\lambda_{\rm Edd}\gtrsim20$. For higher-redshift LRDs, we advocate for rest-near-IR spectroscopic surveys and high-resolution spectra of potential absorption lines as a test of the optically thick atmosphere scenario and as a unique probe of the central engine mass.

\end{abstract}

\keywords{\uat{Active galactic nuclei}{16}, \uat{Radiative transfer}{1335}, \uat{Accretion}{14}}

\section{Introduction} 
\label{sec:intro}
The so-called ``Little Red Dots'' \citep[LRDs,][]{Labbe2023,Kocevski2023,Matthee2024} remain a mystery in extragalactic astronomy. Their broad, luminous emission lines likely indicate a supermassive black hole as the central engine, yet other features set them apart from any previously known population of active galactic nuclei (AGN). Their spectral energy distribution (SED) shows a red rest-optical color and a turnover in the near-infrared (near-IR), resembling a blackbody at $T_{\rm eff}\sim 5000{\rm~K}$ \citep[e.g.,][]{deGraaff2025b,Umeda2025,Inayoshi2025d}. This blue near-IR color, combined with non-detections in the rest-mid-to-far-IR, stringently limits the amount of dust attenuation \citep{Williams2024,Akins2024,Setton2025,Xiao2025,Casey2025,Chen2025}, leaving the red rest-optical unexplained by conventional AGN-galaxy composite models \citep{Baggen2024,Labbe2024,Wang2025,Ma2025}. Furthermore, LRDs do not show the strong X-ray or short-term variability expected from accretion near a black hole \citep{Yue2024,Ananna2024,Maiolino2025,Sacchi2025,Kokubo2024,ZhangZ2025,Burke2025}, which further suggests a physical nature different from typical AGN.

A family of models have invoked spherically symmetric, optically thick atmospheres to explain the observational puzzles (``black hole star'', \citealt{Naidu2025}; gas envelope, \citealt{Kido2025}; spherical super-Eddington accretion, \citealt{Liu2025}; late-stage ``quasi-star'', \citealt{Begelman2025,Santarelli2025}; supermassive star, \citealt{Nandal2026}). The photosphere of the gas naturally produces blackbody-like emission at $T_{\rm eff}\sim5000{\rm~K}$ due to the steep temperature dependence of the hydrogen opacity, which shares a physical origin with the Hayashi track in stellar evolution \citep{Hayashi1961}. The optically thick gas will suppress any X-rays formed near the black hole. The large photosphere radius implies no variability below a timescale measured in decades \citep{ZhangZ2025b}. This scenario, if further confirmed, will redefine our current understanding of accretion physics and black hole evolution. 

To test these models against the rich spectral data, one needs model predictions more detailed than a blackbody. The SEDs of stars are known to deviate from a blackbody, and so do those of LRDs \citep{deGraaff2025b}. It remains unclear under what conditions optically thick gas can reproduce the observed SED shape and various spectral features of LRDs. Some works have used existing stellar spectral libraries to compare to observed LRDs or make model predictions \citep[e.g.,][]{deGraaff2025,Ji2025b,Santarelli2025,Wang2026}, but, as noted in \citet{Liu2025} and \citet{Santarelli2025} from different perspectives, the LRD atmosphere may have densities below those of stars. The emission properties of such low-density atmospheres need modeling to connect theory with observation.

On the observational side, inferences of the black hole mass based on the broad Balmer emission lines are uncertain by orders of magnitude, with ongoing debates over the broadening mechanism of these lines \citep{King2024,Rusakov2026,Naidu2025,Chang2026,Torralba2025,Sneppen2026,Juodzbalis2024,Brazzini2025,Brazzini2026}. Independent constraints on the black hole mass are in high demand. A different methodology is widely adopted in the stellar community: spectral features that originate from the photosphere, e.g., absorption lines, are sensitive tracers of the atmosphere conditions, and fitting theoretical spectra to data constrains parameters such as the effective temperature, surface gravity, and metallicity. As the SED of LRDs resembles stars in the rest-optical to near-IR, a similar approach, i.e., calculating model atmospheres and synthetic spectra from first principles and then comparing them to observations, may provide unique insights to the photosphere properties of LRDs and thus reveal information about their central engines.  

In this work, we construct one-dimensional model atmospheres with radiative transfer and use them to interpret the optical-to-near-IR portion of the LRD spectra. Our approach is analogous to established stellar synthetic spectral libraries \citep[e.g.,][]{Kurucz1979,Gustafsson2008,Hauschildt2025}, but we focus on relatively low gas densities (typically $10^{-12}-10^{-8}{\rm~g~cm^{-3}}$ at the photosphere) and effective temperatures around $5000{\rm~K}$, a parameter space tailored for LRDs but not fully covered in existing libraries. Our synthetic spectra provide a direct testing ground for optically thick envelope models of LRDs and suggest continuum- and absorption line-based observational diagnostics of their physical properties. Our spectral library is publicly available on GitHub\footnote{\url{https://github.com/hanpu-liu/LRD_synthetic_spectral_library}} and is permanently archived at Zenodo\footnote{\url{https://doi.org/10.5281/zenodo.21948309}}, including the spectra, summary statistics, and an example visualization; we welcome requests to share the atmosphere structures or to add calculations for parameters not yet included. 

This paper is organized as follows. In Section~\ref{sec:methods}, we outline our methods and interpretation of the model parameters. Section~\ref{sec:models} describes our model grid. Section~\ref{sec:spectra} elaborates on the continuum and line features as potential diagnostics for LRDs. In Section~\ref{sec:application}, we apply our library to a local LRD and argue for its likely low mass of the central engine. We summarize our work and discuss caveats, implications for modeling, and prospects for observation in Section~\ref{sec:discussion}. Throughout this work, we refer to radiation frequencies and wavelengths in the rest frame unless otherwise noted. We adopt the \citet{PlanckColl2016} cosmological parameters, i.e., $\Omega_m =0.307$, $\Omega_\Lambda =0.693$, $\Omega_b =0.0486$, and $H_0=67.7{\rm~km~s^{-1}~Mpc^{-1}}$, and use the AB magnitude system. 

\section{Methods}
\label{sec:methods}
We use \textsc{tlusty} \citep[Version 208,][]{Hubeny1988,Hubeny2021} to construct atmosphere models for LRDs. The open-source code has been widely used for both stellar \citep[e.g.,][]{Lanz2007,Osorio2020} and AGN accretion disk \citep[e.g.,][]{Hubeny1998,Hubeny2001,Hui2005} spectral modeling.  The physics and numerical algorithms have been detailed in \citet{Hubeny2017b}; here, we summarize its key features relevant to this work and describe our setup.

The stellar mode of \textsc{tlusty} evaluates one-dimensional radiative and hydrostatic equilibrium equations and produces an atmosphere model parameterized by the effective temperature $T_{\rm eff}$ and surface gravity $g$. The code assumes plane-parallel geometry. The validity of this approximation for LRDs is uncertain but plausible, given their large photosphere radii compared to the atmosphere scale height in most of our models, as we shall see in Section~\ref{subsec:gravity}; we will further discuss the gas geometry in Section~\ref{subsec:egg_interpretation}. We assume local thermal equilibrium (LTE), i.e., the ionization stage and level population of all species follow a thermal distribution, fully described by the local gas temperature and density; we will briefly discuss non-LTE effects in Section~\ref{subsec:discussion_model}. We parameterize the gas metallicity $\rm [M/H]$ scaled from the solar abundance in \citet{Asplund2009}. We set a default microturbulence velocity of $\xi_{\rm mtb}=2{\rm~km~s^{-1}}$ but also include $\xi_{\rm mtb}=10{\rm~km~s^{-1}}$ for a subset of our models.  We set a convection mixing length of 1.25 for all of our models. 

Solving the radiative transfer equation requires knowing the frequency-dependent opacity at each location in the atmosphere. We use the opacity table functionality in \textsc{synspec} \citep[Version 54,][]{Hubeny1988,Hubeny2021}, the companion code to \textsc{tlusty}, to pre-compute the absorption opacity in the wavelength range of $900{\rm~\AA}<\lambda<110000{\rm~\AA}$ over 30000 logarithmically equidistant frequency points. We tested using 100000 frequency points and found negligible differences in the resulting atmospheres. We include the line lists of atoms, ions, 19 diatomic molecules (H$_2$, C$_2$, CH, CN, CO, N$_2$, NH, NO, O$_2$, OH, MgH,
MgC, MgN, MgO, SiH, SiC, SiN, SiO, and TiO), and water (${\rm H_2O})$. The atomic and diatomic molecular line lists (except for TiO) are based on Kurucz's list \citep{kurucz_linelist} and updated with data from the National Institute of Standards and Technology database whenever possible, while the TiO and ${\rm H_2O}$ line lists are taken from the ExoMol database \citep{McKemmish2019,Polyansky2018} (with rejection parameters of $-8.5$ and $-9$ respectively); more detailed descriptions can be found in \citet{Hubeny2021}, and the full line lists are available in the \textsc{tlusty} software package indicated in that paper. Molecules are considered both in opacity and in the equation of state. 

The converged model atmosphere from \textsc{tlusty} is then passed to \textsc{synspec} for detailed spectral synthesis. For the Ca H and K resonance lines, we consider partial frequency redistribution (PFR), an effect that is particularly important for low-density atmospheres and strongly impacts the predicted line equivalent widths. We illustrate this point and describe our treatment in Appendix~\ref{app:pfr}. Where our parameter space overlaps with the stellar regime, we compare our spectra with two stellar synthetic spectral libraries, MARCS \citep{Gustafsson2008} and NewEra \citep{Hauschildt2025}, and find good agreement; see Appendix~\ref{app:validation}. 

\subsection{Interpretation of gravity}
\label{subsec:gravity}
In this subsection, we elaborate on the treatment of hydrostatic equilibrium in our models and the corresponding interpretation of the parameter $g$. This is the parameter of highest interest in this work since it is related to the central mass (see below). We emphasize that radiative transfer, which gives rise to spectral features, does not directly know about gravity. The parameter $g$ only indirectly influences the predicted spectrum by changing the photosphere density (see Equation~(\ref{eq:rho_ph}) below) and thus the opacities and optical depths.

In general, at each atmosphere layer, the equation of motion in the radial direction is given by a balance between gravity, dynamical motion of the gas, and accelerations driven by gas pressure $dP_{\rm gas}/dm$ (where $m$ is the column mass) and radiation $g_{\rm rad}$:
\begin{equation}
    g - g_{\rm dyn} - {dP_{\rm gas}\over dm} = g_{\rm rad}={\kappa_F\sigma T_{\rm eff}^4\over c} = 0.40{\rm~cm~s^{-2}}~\kappa_{0.34}T_5^4\,, \label{eq:hydro}
\end{equation}
where $\kappa_F$ is the flux-mean opacity, $\kappa_{0.34}\equiv\kappa_F/(0.34{\rm~cm^2~g^{-1}})$ ($\kappa_{0.34}=1$ for the electron-scattering opacity of fully ionized gas), and $T_5\equiv T_{\rm eff}/(5000{\rm~K})$. Among these terms, $g_{\rm dyn}$ can be estimated by 
\begin{equation}
    g_{\rm dyn} \simeq -{\rm sgn}\left({dv^2\over dR}\right){v^2\over R}\,, \label{eq:g_dyn}
\end{equation}
where $v$ and $R$ are the characteristic velocity and radius of the gas layer in question. We also define
\begin{equation}
    g_{\rm net} \equiv g - g_{\rm dyn}
\end{equation}
as a shorthand. If the gas is bound, including inflows, rotation, turbulence, or failed outflows, we expect $dv^2/dR<0$ and thus $g_{\rm dyn}>0$, which tends to cancel gravity, i.e., $g_{\rm net}<g$; in the limiting case of virial motion, $g_{\rm net}\ll g$.  For outflows that successfully escape the system, we expect $dv^2/dR>0$ and thus $g_{\rm net}>g$.

We note that $g_{\rm net}$ can also be understood as a way of parameterizing the photosphere gas density. Denoting the gas density, Rosseland-mean opacity, and scale height at the photosphere as $\rho_{\rm ph}, \kappa_{\rm ph}$, and $H_{\rm ph}$, we have $\rho_{\rm ph}\kappa_{\rm ph}H_{\rm ph}\sim2/3$ and $dP_{\rm gas}/dm\sim k_BT_{\rm eff}/\mu m_pH_{\rm ph}$, and thus
\begin{equation}
    \rho_{\rm ph} \sim {2\mu m_p (g_{\rm net}-g_{\rm rad}) \over 3k_BT_{\rm eff}\kappa_{\rm ph}} = 2.1\times10^{-9}{\rm~g~cm^{-3}}~\mu_{1.3}g_0T_5^{-1}\kappa_{-3}^{-1}\,, \label{eq:rho_ph}
\end{equation}
\begin{equation}
    H_{\rm ph} \sim {k_BT_{\rm eff}\over\mu m_p(g_{\rm net}-g_{\rm rad})} = 3.2\times10^{11}{\rm~cm}~\mu_{1.3}^{-1}g_0^{-1}T_5\,, \label{eq:h_ph}
\end{equation}
where $\mu$ is the mean molecular weight, $m_p$ is the proton mass, $\mu_{1.3}\equiv\mu/1.3$, $g_0\equiv(g_{\rm net}-g_{\rm rad})/(1{\rm~cm~s^{-2}})$, and $\kappa_{-3}\equiv\kappa_{\rm ph}/(10^{-3}{\rm~cm^2~g^{-1}})$. In the usual case that $\kappa_{\rm ph}$ does not sharply decrease with $\rho_{\rm ph}$, the photosphere gas density will increase with net gravity. We also note that the photosphere radius of LRDs can be estimated as $R_{\rm ph}=\sqrt{L_{\rm atm}/4\pi\sigma T_{\rm eff}^4}=1.5\times10^{16}{\rm~cm}$ for a total atmosphere luminosity $L_{\rm atm}=10^{44}{\rm~erg~s^{-1}}$ and $T_{\rm eff}=5000{\rm~K}$. This is much greater than the photosphere scale height in Equation~(\ref{eq:h_ph}) unless $g_{\rm net}-g_{\rm rad}\lesssim10^{-4}{\rm~cm~s^{-2}}$, which justifies the plane-parallel assumption. Finally, the surface gravity $g$ is related to the total mass (including the black hole and gas) enclosed by the photosphere $M_{\rm tot}$ by
\begin{equation}
    g = {4\pi\sigma T_{\rm eff}^4GM_{\rm tot}\over L_{\rm atm}} = 0.60{\rm~cm~s^{-2}}~M_6T_5^4L_{44}^{-1}\,, \label{eq:gravity}
\end{equation}
where $M_6\equiv M_{\rm tot}/(10^6~M_\odot)$, and $L_{44}\equiv L_{\rm atm}/(10^{44}{\rm~erg~s^{-1}})$.

In the calculations that follow, we neglect dynamical motion since it is not part of the hydrostatic calculations in \textsc{tlusty}.    We reiterate, however, that radiative transfer is separate from dynamics (aside from Doppler shifts which can modify the opacities but are also not included in \textsc{tlusty}).   For a given effective temperature and photosphere density, we expect models with dynamical motions to give similar spectral features and continuum emission as the hydrostatic models considered here. Since $\rho_{\rm ph}$ only depends on $g_{\rm net}$ rather than separately on $g$ and $g_{\rm dyn}$, we can constrain $g_{\rm net}$ using the spectral library while being agnostic about the dynamical environment (i.e., $g$ and $g_{\rm dyn}$ can vary individually subject to the constraint of $g_{\rm net}$). In this sense, our spectral library applies not only to quasi-star-like models but also at least semi-quantitatively to disk models \citep{ZhangC2025,Zwick2025,Chen2026} and quasi-spherical accretion \citep{Liu2025} as long as the energy source is deep within the optically thick atmosphere. In what follows, we use ``$g$'' as a model parameter but interpret it as $g_{\rm net}$.  We can then recover the physical gravity and hence the atmosphere-scale total mass using independent constraints on $g_{\rm dyn}$ (e.g., from line broadening). We will show such analysis later in Section~\ref{sec:application}. 

\subsection{Treatment of super-Eddington scenarios}
\label{subsec:super-Eddington_scenarios}
Low-gravity models potentially become super-Eddington, i.e., with $g<g_{\rm rad}$ in some part of the atmosphere. Note that $g_{\rm rad}$ varies by orders of magnitude within a single atmosphere: at $3000{\rm~K}\lesssim T\lesssim8000{\rm~K}$, the opacity sharply increases with temperature, so the inner, hotter layers of the atmosphere usually have higher $g_{\rm rad}$ and more easily become super-Eddington. This leads to two regimes that differ in their dynamical self-consistency:

A. When $g$ is moderately low, the outer layers still satisfy $g>g_{\rm rad}$ while the inner ones reach $g<g_{\rm rad}$ and thus $dP_{\rm gas}/dm<0$. This will give rise to a gas pressure maximum, $P_{\rm gas,max}$, at $g=g_{\rm rad}$. We modify the hydrostatic assumption by setting $P_{\rm gas}=P_{\rm gas,max}$ in the super-Eddington layers, with the rationale that gas pressure inversion is unstable and that the inner atmosphere may dynamically smear out this inversion layer \citep{Jiang2015}. This dynamical modification does not strongly influence the synthetic spectra as the outer layers, where most spectral features form, remain in hydrostatic equilibrium. In the rest of this work, we refer to models with hydrostatic upper layers as ``hydrostatic'' irrespective of whether they have super-Eddington inner layers. These models are dynamically self-consistent at the photosphere.

B. For even lower $g$, the entire atmosphere becomes super-Eddington. We explore a few cases in this regime by removing radiation acceleration from the hydrostatic equation altogether. In this case, our physical interpretation is no longer an atmosphere model with a hydrostatic structure, but rather an optically thick gas layer with a prescribed (but dynamically unphysical) gas pressure structure given by $dP_{\rm gas}/dm=g$. Equation~(\ref{eq:rho_ph}) is still valid here, with $g_{\rm rad}=0$, and the low $g_{\rm net}$ allows us to obtain low $\rho_{\rm ph}$. In reality, such low-density atmospheres are likely spatially extended with strong gas motion, e.g., in the form of a wind, where our code assumptions break down. The no-$g_{\rm rad}$ models are therefore not dynamically self-consistent atmospheres; they serve only to illustrate the spectral properties of optically thick gas at very low photospheric density, a regime that may be physically realized through dynamics not captured by our calculation. In all figures in this work, we will use filled markers for hydrostatic model points and empty markers for models with $g_{\rm rad}$ turned off to visually distinguish these two regimes. No interpolation across the hydrostatic and no-$g_{\rm rad}$ branches is performed throughout this work.

\section{Model grids}
\label{sec:models}
\begin{figure}
    \centering
    \includegraphics[width=0.99\linewidth]{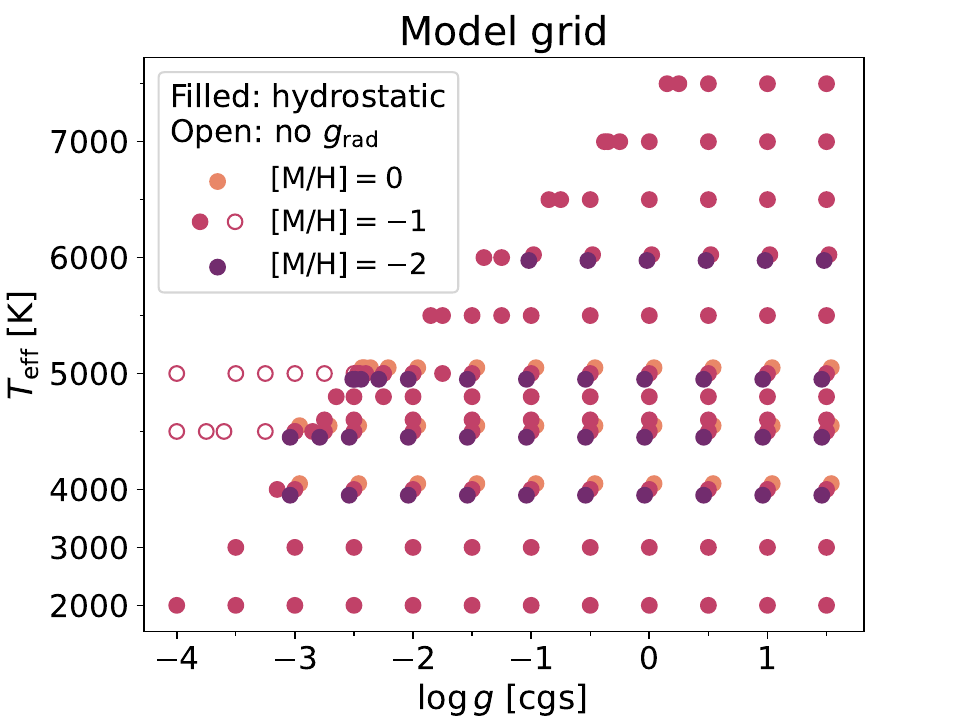}
    \caption{Grid of model parameters covered in this work. Points with multiple metallicity coverage are slightly shifted for visual clarity. The vertical plotting scale is different below $4000{\rm~K}$, where coverage is sparser. Only models with $\xi_{\rm mtb}=2{\rm~km~s^{-1}}$ are shown, but we also calculated hydrostatic models with $\xi_{\rm mtb}=10{\rm~km~s^{-1}}$, $4000{\rm~K}\leq T_{\rm eff}\leq5000{\rm~K}$, and $\rm[M/H]=-1$. Filled markers (``hydrostatic'') are dynamically self-consistent at the photosphere; empty markers (``no $g_{\rm rad}$'') are not dynamically consistent but illustrate spectral properties at very low density. For the full description of each regime, see Section~\ref{subsec:super-Eddington_scenarios}. }
    \label{fig:grid}
\end{figure}

Figure~\ref{fig:grid} outlines the parameter space explored in this work in the dimensions of $T_{\rm eff}, \log g$, and $\rm [M/H]$. The model grid spans a range of $2000{\rm~K}\leq T_{\rm eff}\leq 7500{\rm~K}$ and is densest at $T_{\rm eff}=4500-5000{\rm~K}$, motivated by LRDs clustering at this optical color temperature \citep{deGraaff2025b}. Our calculations mainly cover $\rm [M/H]=-1$; we will show later that different metallicities significantly change metal absorption features but only mildly influence the spectral continuum. While Figure~\ref{fig:grid} only shows models with $\xi_{\rm mtb}=2{\rm~km~s^{-1}}$, we also include $\xi_{\rm mtb}=10{\rm~km~s^{-1}}$ for hydrostatic models with $4000{\rm~K}\leq T_{\rm eff}\leq 5000{\rm~K}$ and $[\rm M/H]=-1$; we will show later that microturbulence tends to redden the UV to blue-optical portion of the SED by metal line-blocking but has negligible impact on the continuum at longer wavelengths. As for the range of $g$, we reach a maximum of $\log g=1.5$ (in $\rm cm~s^{-2}$), which corresponds to $M_{\rm tot}\sim10^8~M_\odot$ according to Equation~(\ref{eq:gravity}) if assuming radial hydrostatic equilibrium. For each $T_{\rm eff}$, we decrease $\log g$ until the upper atmosphere becomes super-Eddington (Section~\ref{subsec:gravity}) or until $\log g=-4$. This gives the left boundary of the filled markers in Figure~\ref{fig:grid}. For $T_{\rm eff}=4500{\rm~K}$ and 5000~K, we further explore the no-$g_{\rm rad}$ models (i.e., sacrificing dynamical consistency) and reach $\log g=-4$, indicated by the empty markers.

\begin{figure}
    \centering
    \includegraphics[width=0.99\linewidth]{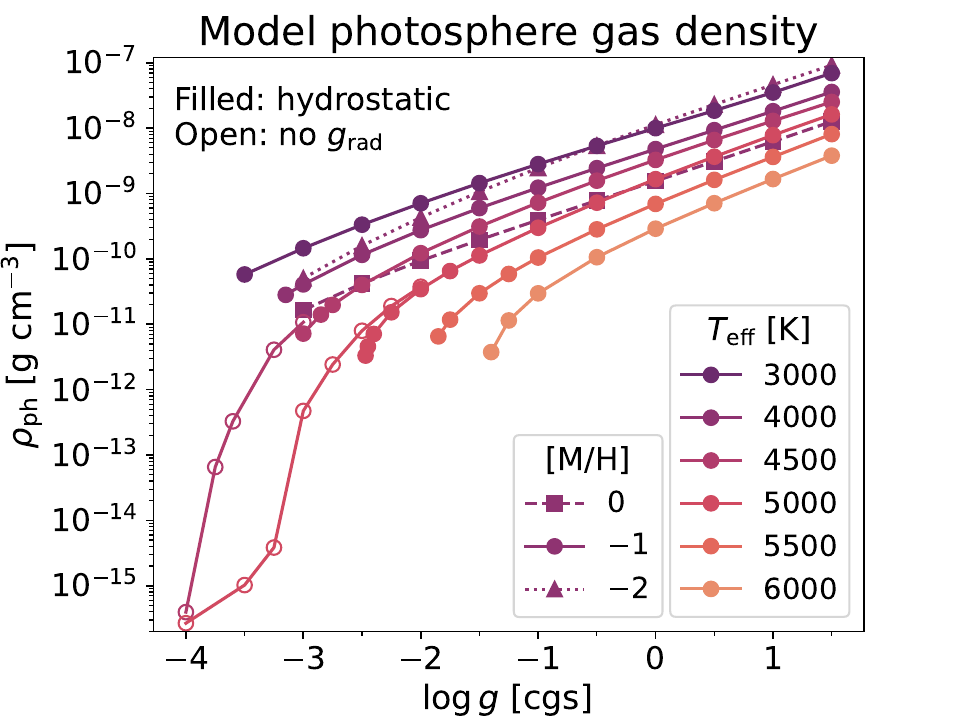}
    \caption{Photosphere gas density $\rho_{\rm ph}$ as a function of gravity. Multiple metallicities are shown only for $T_{\rm eff}=4000{\rm~K}$ for visual clarity. Only models with $\xi_{\rm mtb}=2{\rm~km~s^{-1}}$ are shown; results are similar for $\xi_{\rm mtb}=10{\rm~km~s^{-1}}$. Filled markers (``hydrostatic'') are dynamically self-consistent at the photosphere; empty markers (``no $g_{\rm rad}$'') are not dynamically consistent but illustrate spectral properties at very low density (see Section~\ref{subsec:super-Eddington_scenarios}). For a given $T_{\rm eff}$ and $\rm [M/H]$, $\rho_{\rm ph}$ increases with $\log g$. }
    \label{fig:photosphere_density}
\end{figure}

The low end of the gravity range, which implies $M_{\rm tot}\sim10^3~M_\odot$ by Equation~(\ref{eq:gravity}), seems implausible if LRDs are in radial hydrostatic equilibrium.   However, the photosphere densities that these $\log g$ values represent are possible in alternative scenarios. Figure~\ref{fig:photosphere_density} shows the photosphere gas density, $\rho_{\rm ph}$, as a function of $\log g$. We measure $\rho_{\rm ph}$ from the converged model atmospheres at the location where the Rosseland-mean optical depth is equal to 2/3. For a given $T_{\rm eff}$ and $\rm [M/H]$, $\rho_{\rm ph}$ increases with $\log g$, as expected from Equation~(\ref{eq:rho_ph}). For $T_{\rm eff}=5000{\rm~K}$, a factor of 10 higher or lower metallicity from the fiducial value of $\rm [M/H]=-1$ introduces a change of $\rho_{\rm ph}$ by a factor $<3$; the same is true for $T_{\rm eff}=4000{\rm~K}$ and $6000{\rm~K}$, although not shown in the figure. Microturbulence does not strongly influence the photosphere density, with $\rho_{\rm ph}$ differing by a factor $<1.5$ between $\xi_{\rm mtb}=2{\rm~km~s^{-1}}$ and $10{\rm~km~s^{-1}}$. Most radially hydrostatic models have $\rho_{\rm ph}\gtrsim10^{-11}{\rm~g~cm^{-3}}$. However, the disk models in \citet{Zwick2025,Chen2026} and the super-Eddington accretion model in \citet{Liu2025} do reach densities of the order $10^{-12}{\rm~g~cm^{-3}}$ or lower even with high central masses. For these scenarios, Figure~\ref{fig:photosphere_density} is the tool to convert ``$\log g$'' as a density parameter to the physically meaningful $\rho_{\rm ph}$. We note that our grid reaches much lower $\log g$ than stellar synthetic spectral libraries, which typically cover $\log g\geq0$. 

In Figure~\ref{fig:photosphere_density}, at $T_{\rm eff}=5000{\rm~K}$ and $\rm [M/H]=-1$, the hydrostatic branch (where we account for $g_{\rm rad}$) and the branch without $g_{\rm rad}$ agree at $\log g\gtrsim-2.25$ and only diverge below this. Meanwhile, we cannot find a hydrostatic solution at $\log g\leq-2.48$, indicating that, at the photosphere, $\log g_{\rm rad}\approx-2.48$. This implies that radiation acceleration only strongly influences $\rho_{\rm ph}$ when $g$ approaches $g_{\rm rad}$ at the photosphere, as expected from Equation~(\ref{eq:rho_ph}). We note that $\log g_{\rm rad}\approx-2.48$ is much lower than the radiative acceleration of the inner, fully ionized layers (Equation~(\ref{eq:hydro})) due to the much lower opacity at the photosphere.

\section{Synthetic spectra}
\label{sec:spectra}
In this section, we investigate several continuum and line features of our synthetic spectra, in particular their parametric dependence and their observational utility in diagnosing the physical conditions of LRDs.

\begin{figure*}
    \centering
    \includegraphics[width=0.99\linewidth]{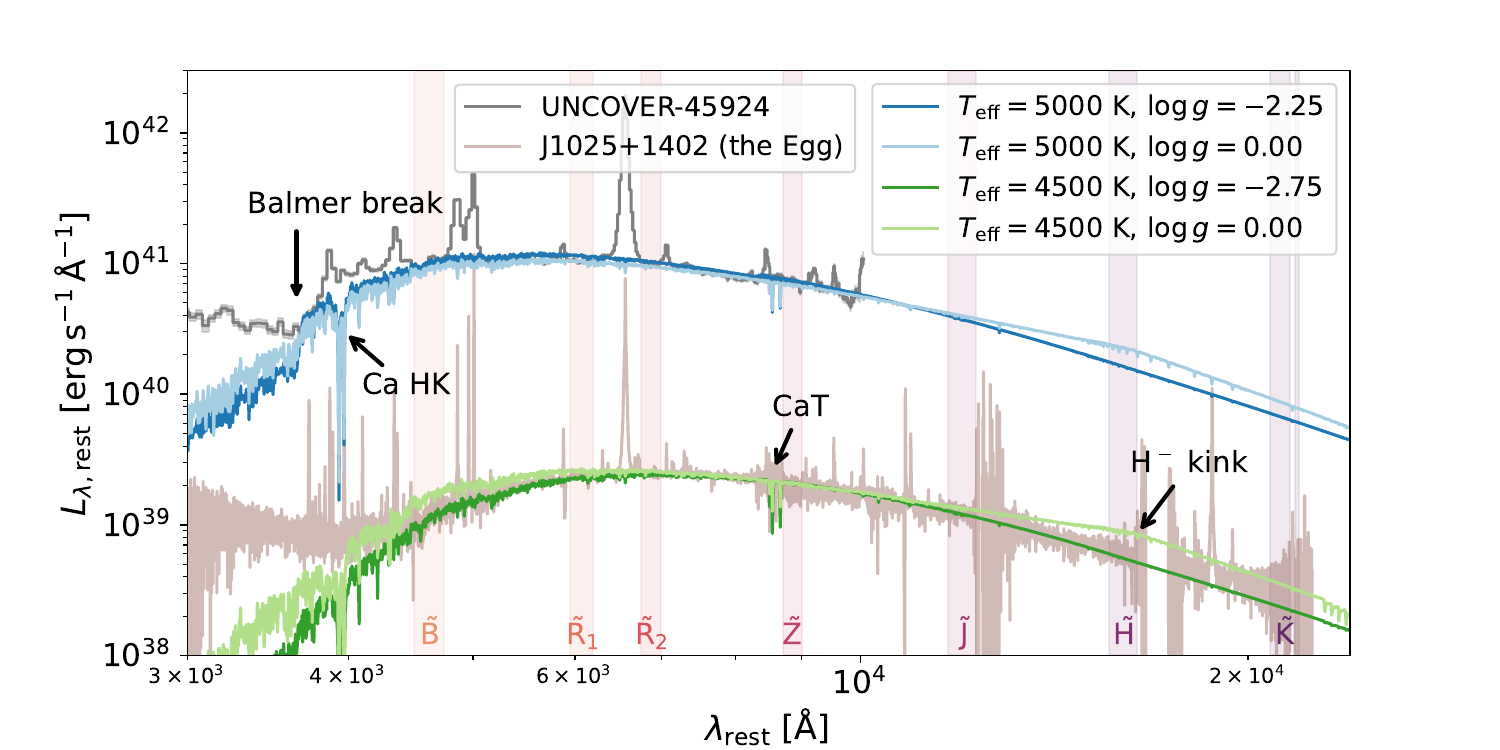}
    \caption{Overview of synthetic spectra and the spectral features to be discussed. Two models at $T_{\rm eff}=5000{\rm~K}$ broadly agree with the spectrum of UNCOVER-45924 \citep{Labbe2024}, but their different $\log g$ values give different Balmer break and Ca HK absorption strengths. Two models at $T_{\rm eff}=5000{\rm~K}$ broadly agree with the spectrum of J1025-1402 \citep[nicknamed \egg,][]{Lin2025b}, but the low-gravity one gives a narrower optical-to-near-IR SED, stronger \ion{Ca}{2} triplet (CaT) absorption, and does not show a continuum turnover at $\lambda_{\rm rest}\sim1.6{\rm~\mu m}$ (the ``$\rm H^-$ kink''). The low-gravity model is preferred for \egg\ if we account for the additional galaxy and dust emission (Section~\ref{sec:application}). All four models have $\rm[M/H]=-1$ and $\xi_{\rm mtb}=2{\rm~km~s^{-1}}$. The wavelength ranges of our synthetic photometry bands to be used later, from $\tilde{\rm B}$ to $\tilde{\rm K}$, are shaded in color. }
    \label{fig:overview}
\end{figure*}

Figure~\ref{fig:overview} gives an overview of the synthetic spectra. We can qualitatively match the optical-to-near-IR spectra of LRDs such as UNCOVER-45924 \citep{Labbe2024} or J1025+1402 \citep[nicknamed \egg,][]{Lin2025b} using a single $T_{\rm eff}$, in agreement with the findings in \citet{deGraaff2025b,Umeda2025} that the SEDs of LRDs generally appear similar to a blackbody at this wavelength range. This supports the interpretation of LRDs as an optically thick emission medium with a narrow range of effective temperature in which the details of the inner energy source are largely hidden. However, beyond the zeroth-order agreement, models with different $\log g$ (atmosphere density) show distinct features in the continuum and lines. Figure~\ref{fig:overview} provides a roadmap to the following subsections, where we discuss the parametric dependence of various features.

We note that \egg\ \citep[$z_{\rm spec}=0.1007$,][]{Lin2025b,Ji2025b} is among the lowest-redshift LRDs known so far. It has the rare combined advantage of broad spectroscopic wavelength coverage and high spectral resolution. This object will be the main point of comparison in this section.

To quantify the SED profile, we design seven medium-width photometric bands with top-hat filters: $\tilde{\rm B}$ ($4500-4750\rm~\AA$), $\tilde{\rm R}_1$ ($5950-6200\rm~\AA$), $\tilde{\rm R}_2$ ($6750-7000\rm~\AA$), $\tilde{\rm Z}$ ($8700-9000\rm~\AA$), $\tilde{\rm J}$ ($11700-12300\rm~\AA$), $\tilde{\rm H}$ ($15600-16400\rm~\AA$), and $\tilde{\rm K}$ ($20800-21550, 21750-21900\rm~\AA$), as indicated in Figure~\ref{fig:overview}. We use the tilde symbol to avoid confusion with the standard broadband filters. The wavelengths are chosen to avoid strong emission or absorption lines. We measure the synthetic photometry for both our model spectra and real data in the following subsections. 

\subsection{Optical SED}

\begin{figure*}
    \centering
    \begin{minipage}{0.45\textwidth}
        \centering
        \includegraphics[width=0.99\linewidth]{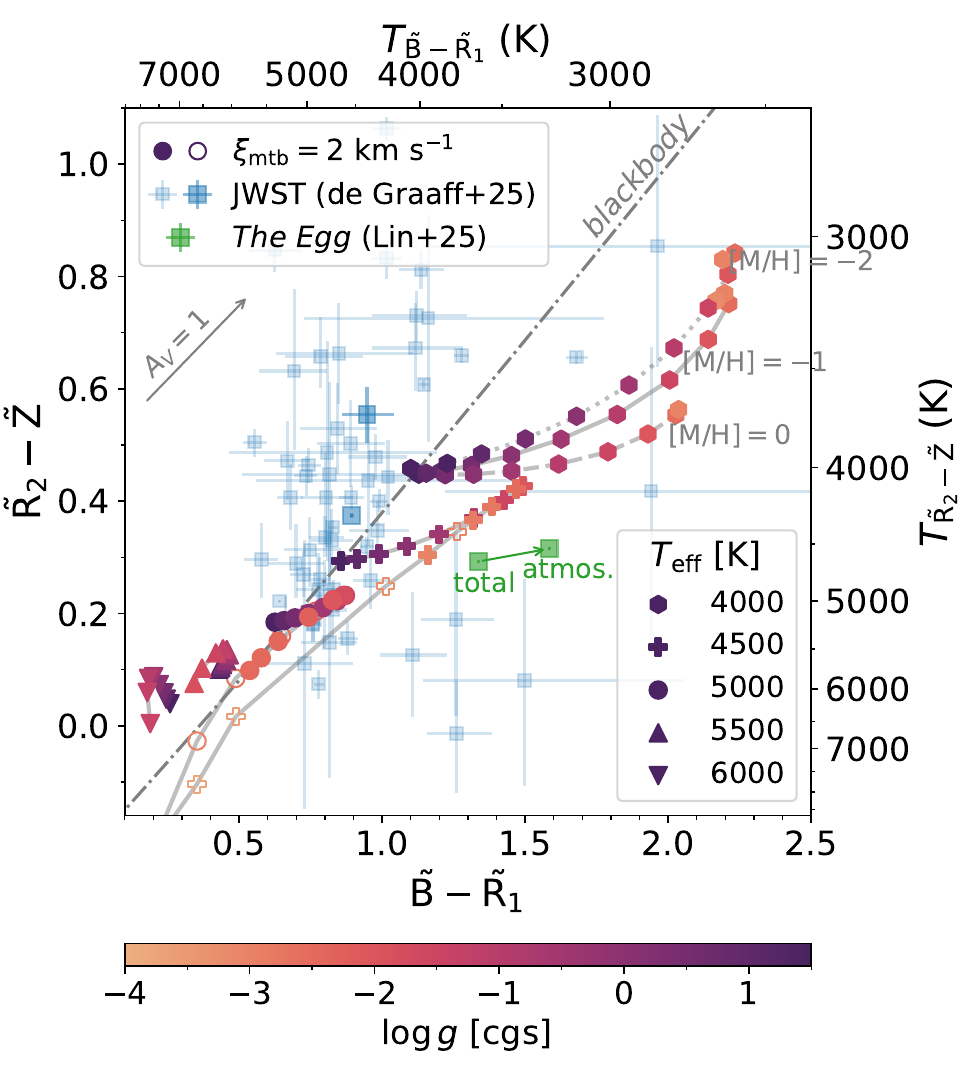}
    \end{minipage}
    \begin{minipage}{0.45\textwidth}
        \centering
        \includegraphics[width=0.99\linewidth]{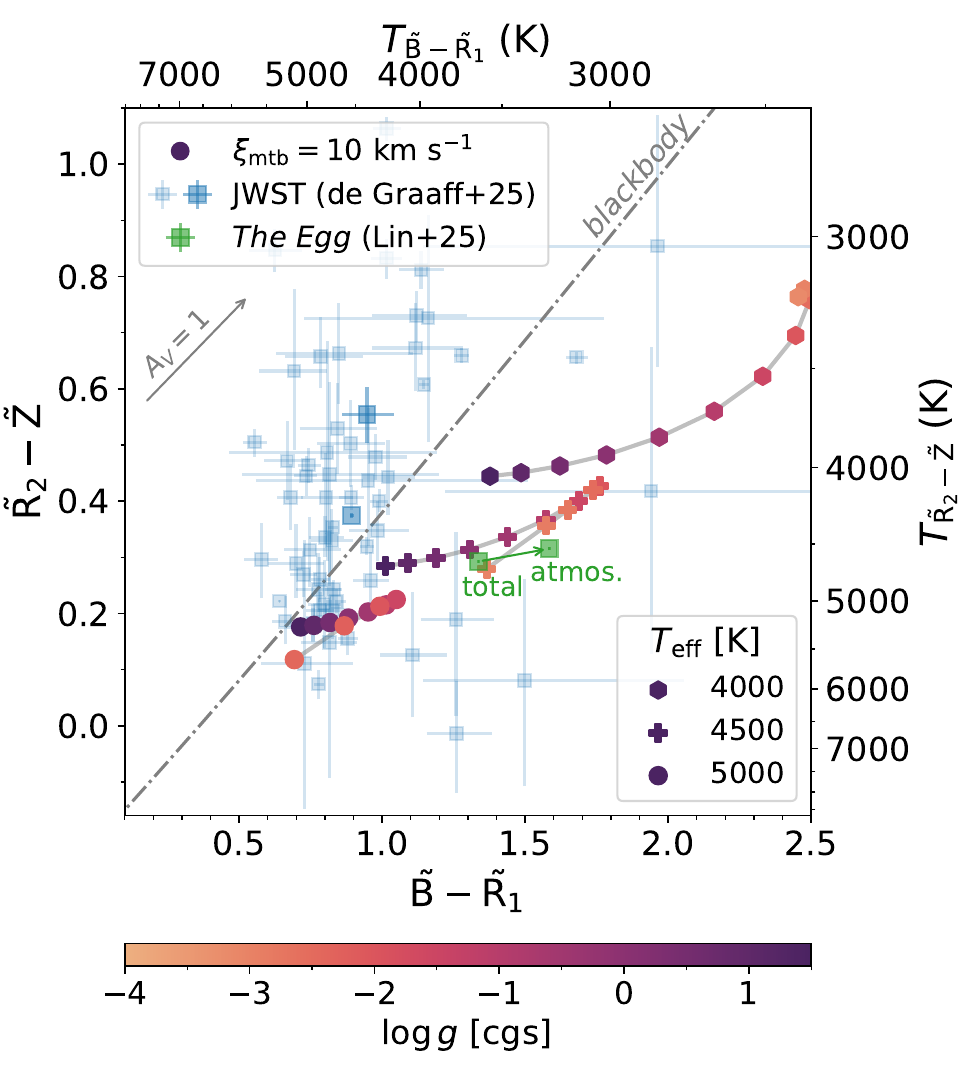}
    \end{minipage}
    \caption{Color-color diagram of synthetic spectra (left: $\xi_{\rm mtb}=2{\rm~km~s^{-1}}$; right: $\xi_{\rm mtb}=10{\rm~km~s^{-1}}$) and observed LRDs. $\rm \tilde{B}-\tilde{R}_1$ and $\rm \tilde{R}_2-\tilde{Z}$ (see Figure~\ref{fig:overview} and text for band definitions) capture the optical and near-IR slopes. Axes on the top and right give the color temperatures. Points below the dash-dotted ``blackbody'' line of $T_{\rm \tilde{B}-\tilde{R}_1}=T_{\rm \tilde{R}_2-\tilde{Z}}$ have SEDs narrower than a blackbody, i.e., with relatively red optical colors but blue near-IR colors. Gray vector indicates the change of colors by an SMC dust extinction of $A_V=1$. Model points are color-coded by $\log g$ (which proxies the photospheric density); those with the same $T_{\rm eff}$ and [M/H] are connected by gray lines. Three metallicity values are shown for $T_{\rm eff}=4000{\rm~K}$ and $\xi_{\rm mtb}=2{\rm~km~s^{-1}}$, while only $\rm[M/H]=-1$ is shown for other cases. Empty markers indicate no-$g_{\rm rad}$ models. The JWST LRD samples in \citet{deGraaff2025b} and \egg\ \citep{Lin2025b} are shown in comparison. JWST data points measure the original spectrum; for \egg, the ``total'' point measures the original spectrum and the ``atmos.'' point measures the spectrum subtracted by a best-fit young galaxy model (Section~\ref{sec:application}). For cool models with $T_{\rm eff}<5000{\rm~K}$, low-density (low-$\log g$) atmospheres may give SEDs significantly narrower than a blackbody. }
    \label{fig:color_color}
\end{figure*}

We first focus on the overall shapes of our synthetic spectra. Recent demographic studies have quantified the optical-to-near-IR SED of LRDs in several ways, including fitting to a modified blackbody \citep{deGraaff2025b} and to polynomials \citep{Barro2026}. We take a different approach and use colors to capture the optical and near-IR slopes. Two advantages of this representation are firstly, that the color measurement does not suffer from fitting systematics due to the inhomogeneous wavelength coverage of real data, and secondly, that our band coverage is designed to be relatively line-free and thus does not require the sometimes nontrivial task of masking lines. 

We first study the wavelength range of approximately $4500-9000{\rm~\AA}$, characterized by colors $\rm\tilde{B}-\tilde{R}_1$ and $\rm \tilde{R}_2-\tilde{Z}$. Figure~\ref{fig:color_color} shows the color-color diagram of our synthetic spectra. To aid interpretation, we show the color temperatures on the top and right axes. For example, in the left panel, the point at $T_{\rm eff}=4500{\rm~K}, \log g=-2.0, \xi_{\rm mtb}=2{\rm~km~s^{-1}}$ has $\rm \tilde{B}-\tilde{R}_1=1.49$ and $\rm \tilde{R}_2-\tilde{Z}=0.43$; the corresponding color temperatures are $T_{\rm \tilde{B}-\tilde{R}_1}=3.4\times10^3{\rm~K}$ and $T_{\rm \tilde{R}_2-\tilde{Z}}=4.1\times10^3{\rm~K}$, i.e., a blackbody at $3.4\times10^3{\rm~K}$ or $4.1\times10^3{\rm~K}$ would have the same $\rm \tilde{B}-\tilde{R}_1$ or $\rm \tilde{R}_1-\tilde{Z}$ color as this model. Since $T_{\rm \tilde{B}-\tilde{R}_1}<T_{\rm \tilde{R}_1-\tilde{Z}}$, the SED of this model is narrower than a blackbody in the sense of having a relatively red optical color but a blue near-IR color. More generally, all points below the diagonal ``blackbody'' line, which delineates $T_{\rm \tilde{B}-\tilde{R}_1}=T_{\rm \tilde{R}_2-\tilde{Z}}$ (on which all perfect blackbodies would lie), have SEDs narrower than a blackbody, and vice versa. For $T_{\rm eff}=4000{\rm~K}$, a higher metallicity gives somewhat bluer colors at a given $\log g$ due to the backwarming effect of metal lines \citep[e.g.,][]{Gustafsson1975}, but this trend weakens at higher temperature, with a color change of no more than 0.03 mag between ${\rm[M/H]=0}$ and ${\rm[M/H]=-1}$ for $T_{\rm eff}=5000{\rm~K}$. We do not show multiple metallicity values for higher $T_{\rm eff}$ for visual clarity.

Among the model points, low $T_{\rm eff}$ values unsurprisingly tend to give red optical and near-IR colors. Notably, the color temperatures do not always follow the effective temperature. When $T_{\rm eff}<5000{\rm~K}$, while models at $\log g\sim1.5$ have $T_{\rm \tilde{B}-\tilde{R}_1}\approx T_{\rm \tilde{R}_2-\tilde{Z}}\approx T_{\rm eff}$, lowering $\log g$ can reduce the color temperatures by more than $1000{\rm~K}$. $T_{\rm \tilde{B}-\tilde{R}_1}$ is reduced more than $T_{\rm \tilde{R}_2-\tilde{Z}}$, and the SED becomes narrower than a blackbody at low $\log g$. This is because the absorption opacity at near-IR wavelengths, dominated by $\rm H^{-}$, scales more steeply with density than the absorption opacity in the optical, which is dominated by metal lines. At low photosphere density, the optical opacity becomes comparatively larger than the near-IR opacity, and thus the optical continuum is produced further out in colder atmosphere layers. At very low $\log g$ (e.g., $\log g < -2$ for $T_{\rm eff}=4500{\rm~K}$), the colors start to move bluer at lower $\log g$ due to the increasing importance of the scattering opacity, which makes the continuum form at deeper, hotter layers than the Rosseland-mean photosphere. As for microturbulence, a comparison between the two panels in Figure~\ref{fig:color_color} shows that a high $\xi_{\rm mtb}$ reddens the $\rm \tilde{B}-\tilde{R}_1$ color and thus makes the SED narrower. This is due to metal absorption lines blending into a pseudo-continuum at blue optical wavelengths, and a high $\xi_{\rm mtb}$ promotes blending by boosting the line equivalent widths. 

In comparison, we plot in Figure~\ref{fig:color_color} the colors of JWST spectroscopically confirmed LRD samples from \citet{deGraaff2025b}. We only include LRDs at redshifts $z\lesssim5$, whose $\tilde{\rm Z}$ bands are covered by JWST/NIRSpec. We calculate the colors using the PRISM spectra from the public Dawn JWST Archive, with the 1-$\sigma$ uncertainties estimated from 1000 random draws of the error spectra. We similarly calculate the colors of \egg\ using the spectral data reported in \citet{Lin2025b}. 

We first note that a fraction of LRDs, including \egg, show SEDs narrower than a blackbody; see also \citet{Lin2025b,deGraaff2025b}. This will be difficult to explain if one only assumes blackbody radiation. Dust extinction will not help: the vector of dust extinction (``$A_V=1$'', assuming a Small Magellanic (SMC)-averaged dust law, \citealt{Gordon2024}) is almost parallel to the ``blackbody'' line, which means that a dust-extincted blackbody is hardly narrower than a true blackbody in this color-color space. If we were to plot a line of blackbodies reddened by a generous $A_V=3$, it would only be horizontally displaced from the ``blackbody'' line by $<0.2$ mag. Moreover, there is mounting evidence for a physically distinct UV component of LRDs in addition to the atmosphere considered here \citep{Rinaldi2024,Chen2025a,Chen2025b,Barro2026,Baggen2026}, likely due to the host galaxy or a blue companion. Accounting for this component will intensify the narrowness. For example, if we subtract a young galaxy model from \egg\ (to be described in Section~\ref{sec:application}), its location in color-color space will depart farther from the ``blackbody'' line (Figure~\ref{fig:color_color}) because the galaxy contributes significantly to the $\tilde{\rm B}$ band but much less so at longer wavelengths.  Our synthetic spectra naturally reproduce narrow SEDs because of the radiative transfer effects described above. 

On the other hand, many JWST data points are clustered at $T_{\rm \tilde{B}-\tilde{R}_1}\sim4500-5000{\rm~K}$, with comparable or slightly lower $T_{\rm \tilde{R}_2-\tilde{Z}}$, implying an SED shape comparable to or mildly broader than a blackbody. This agrees with the findings of \citet{deGraaff2025b}. These samples appear to lie beyond our model points or suggest high $\log g$. An observational complication is that the atmospheres may have redder $\rm \tilde{B}-\tilde{R}_1$ colors after subtracting the UV component. As a crude estimate, if the atmosphere contributes 60\% and 80\% of the total flux in the $\tilde{\rm B}$ and $\tilde{\rm R}_1$ bands respectively \citep[Figure~12 in][]{Sun2026}, then the location of the intrinsic atmosphere in the color-color space will shift rightwards by $\Delta(\rm\tilde{B}-\tilde{R}_1)=-2.5\log(60\%/80\%)=0.3$ mag.  Even after this decomposition, the colors of some observed LRDs will still deviate from the models presented here. This indicates that a single optically thick atmosphere as considered in this work is too simplistic to explain the optical-to-near-IR colors of many LRDs.  A range of atmosphere conditions in $T_{\rm eff}$ and/or $\log g$ may account for a broad SED, in which case our models can provide a basis for constructing multi-component spectral fits. Indeed, a range of $T_{\rm eff}$ or $\log g$ is required to explain certain spectral features, e.g., the water absorption in \citet{Wang2026} and the smooth Balmer limit in \citet{Hviding2026}. We caution that an alternative explanation for the observed broadness in SED is that the optical-to-near-IR continuum emission from LRDs is not fully optically thick \citep[as in the ``cocoon'' scenario,][]{Sneppen2026}, in which case our models will cease to be applicable.  

\subsection{Near-IR SED}
\label{subsec:near_IR_SED}
We have found in the previous section that our models naturally explain the optical SED of \egg. Its rest-near-IR spectral coverage motivates us to investigate the SED at longer wavelengths. Moreover, inferring parameters such as $\log g$ from the optical alone is challenging: even without the uncertainty of host decomposition, degeneracy is severe as the colors non-monotonically change with $\log g$ and are sensitive to $\xi_{\rm mtb}$. However, the near-IR SED can serve as an effective probe for $\log g$, as we show below. 

\begin{figure}
    \centering
    \includegraphics[width=0.99\linewidth]{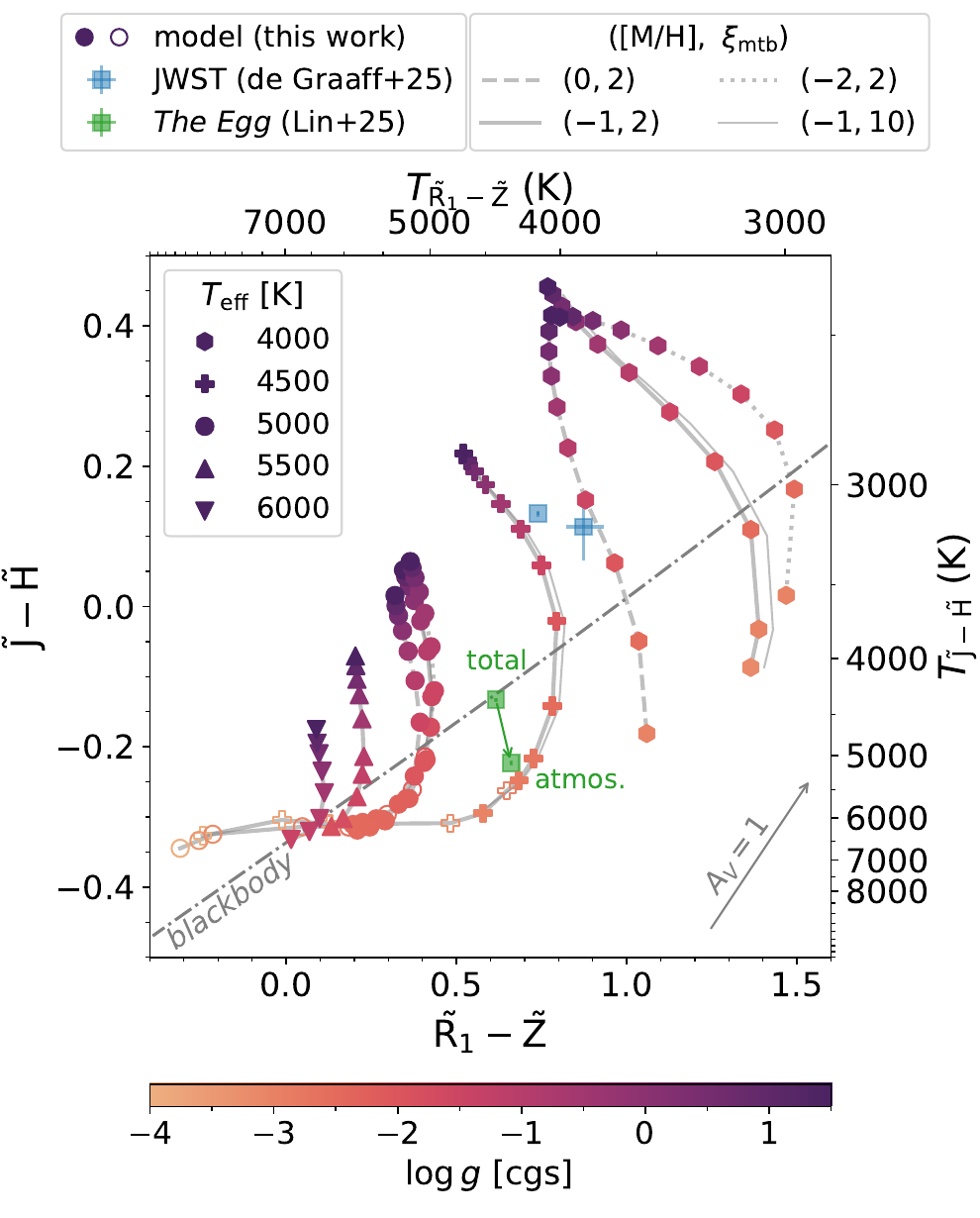}
    \caption{Similar to Figure~\ref{fig:color_color}, but for longer wavelengths, i.e., $\rm \tilde{R}_1-\tilde{Z}$ vs. $\rm \tilde{J}-\tilde{H}$. Three metallicity values are shown for $T_{\rm eff}=4000,5000{\rm~K}$ and $\xi_{\rm mtb}=2{\rm~km~s^{-1}}$, while only $\rm[M/H]=-1$ is shown for other cases. Microturbulence $\xi_{\rm mtb}=10{\rm~km~s^{-1}}$ is shown as thin solid lines for $T_{\rm eff}=4000, 4500, 5000{\rm~K}$ and $\rm[M/H]=-1$ but without markers for visual clarity. The JWST LRD samples in \citet{deGraaff2025b} and \egg\ \citep{Lin2025b} are shown in comparison. JWST data points measure the original spectrum; for \egg, the ``total'' point measures the original spectrum and the ``atmos.'' point measures the spectrum subtracted by a best-fit young galaxy model dominating in the UV and a warm dust emission model dominating in the mid-IR (Section~\ref{sec:application}). Model colors at long wavelengths strongly depend on photospheric density (proxied by $\log g$) and are insensitive to $\xi_{\rm mtb}$; the models suggest at least $\log g<-2$ for \egg.   }
    \label{fig:color_color_IR}
\end{figure}

Figure~\ref{fig:color_color_IR} presents the color-color space at longer wavelengths, around $6000-16000{\rm~\AA}$.  Similarly to Figure~\ref{fig:color_color}, the color temperatures of the models can deviate drastically from $T_{\rm eff}$, and smaller $\log g$ tends to give narrower SEDs (at least for hydrostatic cases). However, we highlight two important differences from the optical trends. First, for a given $T_{\rm eff}$ and $\rm[M/H]$, there is an almost one-to-one mapping from $\log g$ to $\rm \tilde{J}-\tilde{H}$ over a wide range of $\log g$. This trend is driven by the decreasing importance of the $\rm H^-$ opacity relative to the electron scattering opacity with decreasing atmosphere density. Second, this mapping is insensitive to $\xi_{\rm mtb}$: the thin solid gray curves in Figure~\ref{fig:color_color_IR} almost overlap with the thick solid curves. This is because, unlike in the blue optical, atomic lines at this wavelength are too sparse to blend into a pseudo-continuum and affect the colors. These two advantages will allow us to put strong constraints on $\log g$ if the near-IR SED is available. 

We now investigate what this implies for \egg. In Figure~\ref{fig:color_color_IR}, we show the measured near-IR colors from the observed spectrum (the ``total'' point). The point is located very close to the ``blackbody'' line. Meanwhile, on its left and right, the models with $4000{\rm~K}\leq T_{\rm eff}\leq5000{\rm~K}$ and $\rm[M/H]=-1$ cross the ``blackbody'' line at $-2.5<\log g<-2.0$. This suggests that $\log g<-2.0$ for \egg. This can be even more stringent. Based on mid-IR photometry, \citet{Lin2025b} found evidence for warm dust emission. Its high-frequency tail will enter the $\tilde{\rm H}$ band and make $\rm \tilde{J}-\tilde{H}$ appear redder than the intrinsic atmosphere. After subtracting the aforementioned galaxy component (which contributes $7\%$ and $3\%$ to the $\tilde{\rm R}_1$ and $\tilde{\rm Z}$ band fluxes) and the dust emission component (which we model in Section~\ref{sec:application}, with results in agreement with \citealt{Lin2025b}), we find that the intrinsic colors of \egg\ (the ``atmos.'' point in Figure~\ref{fig:color_color_IR}) are close to $T_{\rm eff}=4500{\rm~K}$ and $\log g=-2.85$. The ``atmos.'' point does not correct for potential dust extinction (see the ``$A_V=1$'' vector; not to be confused with dust emission), which would imply yet a narrower SED and a lower $\log g$. In conclusion, the strong, robust relation between $\log g$ and near-IR colors allows us to constrain at least $\log g<-2.0$ for \egg\ or even approaching $\log g\approx-3$ when considering potential dust emission. 

At higher redshifts, two LRDs in \citet{deGraaff2025b} have JWST/NIRSpec coverage of the $\tilde{\rm H}$ band. They are \textit{the Rosetta Stone} \citep{Juodzbalis2024} and GN-28074, highlighted with big blue squares in Figures~\ref{fig:color_color} and \ref{fig:color_color_IR}. \textit{The Rosetta Stone} has smaller error bars than GN-28074. Detailed modeling of these two objects is beyond the scope of this paper, but we note that \textit{the Rosetta Stone} is closest to the model point at $T_{\rm eff}=4500{\rm~K}$ and $\log g=-1$, which is an upper limit since corrections for dust emission and/or extinction will imply even lower $\log g$. We will briefly discuss this object in Section~\ref{subsec:discussion_observation}. 

Recently, \citet{Sneppen2026} reported cocoon models that partially reprocess an incident ionizing spectrum and reproduce many spectral features of LRDs. They predicted a near-IR spectral slope of $F_\lambda\propto\lambda^{-2}$, in better agreement with some LRDs than the Rayleigh-Jeans tail of a blackbody, $F_\lambda\propto\lambda^{-4}$. However, our optically thick atmosphere models are also redder than $F_{\rm \lambda}\propto\lambda^{-4}$ (which has a very blue color of $\rm \tilde{J}-\tilde{H}=-0.62$) and can predict a slope of $F_\lambda\propto\lambda^{-2}$ (corresponding to $\rm \tilde{J}-\tilde{H}=0.0$) for some combinations of $T_{\rm eff}$ and $\log g$.  Therefore, the relatively red observed near-IR slopes may disfavor a simple Rayleigh-Jeans tail but do not rule out an optically thick atmosphere with more realistic radiative transfer. We will discuss how the cocoon and atmosphere scenarios may be distinguished in Section~\ref{subsec:discussion_observation}.

\subsection{The $\rm H^{-}$ kink in the near-IR}
\label{subsec:kink}
\begin{figure}
    \centering
    \includegraphics[width=0.99\linewidth]{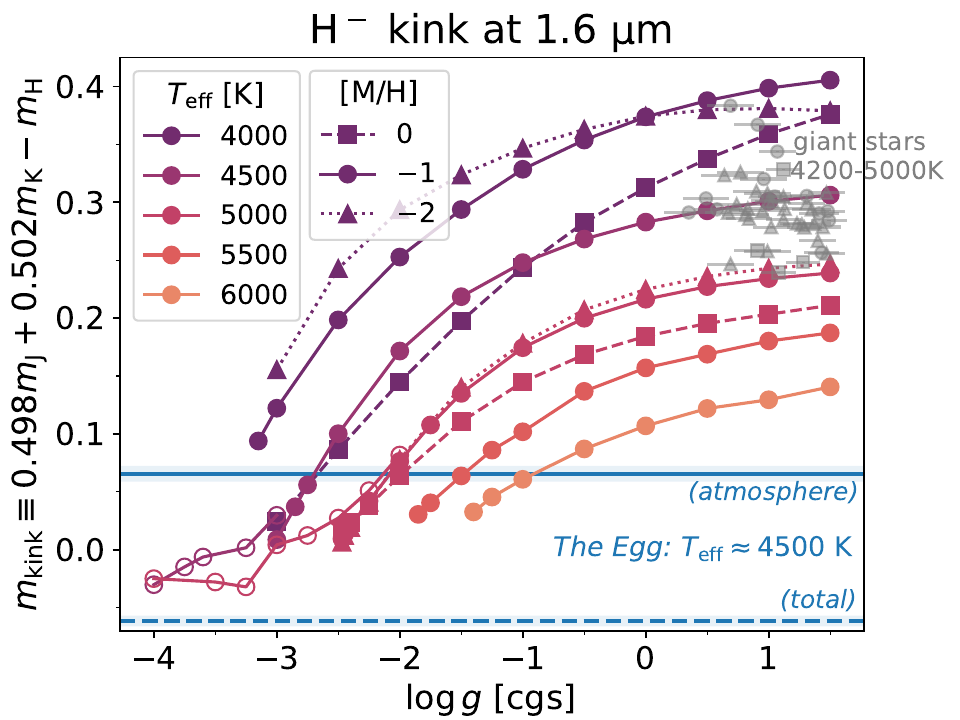}
    \caption{The continuum kink strength at $\lambda_{\rm rest}\sim1.6{\rm~\mu m}$ due to $\rm H^{-}$, defined in Equation~(\ref{eq:kink}). High values of the kink index $m_{\rm kink}$ corresponds to strong kinks. All models shown here have $\xi_{\rm mtb}=2{\rm~km~s^{-1}}$ ($m_{\rm kink}$ is insensitive to $\xi_{\rm mtb}$). The index of \egg, measured from the observed spectrum directly (``total'') and from the spectrum with dust emission subtracted (``atmosphere''), are shown with horizontal dashed and solid blue lines, with the shade representing formal $1\sigma$ errors. Gray points show our kink measurements using the spectra of giant stars with $4200{\rm~K}\leq T_{\rm eff}\leq5000{\rm~K}$ and $\log g\leq1.5$ (spectra from \citealt{Gonneau2020}; atmosphere parameter values and uncertainties from \citealt{Arentsen2019}), which are grouped into three metallicity bins, $\rm -0.5\leq[Fe/H]<0.1$, $\rm -1.5\leq[Fe/H]<-0.5$, and $\rm -2.6<[Fe/H]<-1.5$, indicated by square, circle, and triangular markers respectively. Compared to stars at similar $T_{\rm eff}$, \egg\ ($T_{\rm eff}\approx4500{\rm~K}$) shows a much weaker kink, which suggests a very low atmosphere density, with $\log g\approx-3$, i.e., $\rho_{\rm ph} \approx 10^{-11} {\rm~g~cm^{-3}}$ (Figure~\ref{fig:photosphere_density}).}
    \label{fig:nearIR_triangle}
\end{figure}

The near-IR colors described above are related to a spectral feature at slightly longer wavelengths. In Figure~\ref{fig:overview}, the synthetic spectra at $\log g=0.00$ show a turnover at $\lambda\sim1.6{\rm~\mu m}$. We refer to this continuum feature as the ``$\rm H^-$ kink'': the wavelength corresponds to the ionization limit of the $\rm H^-$ ion, whose opacity increases at both shorter and longer wavelengths due to bound-free and free-free transitions, respectively \citep[ch.~4-4,][]{Mihalas1978}. When $\rm H^-$ is the dominant opacity source, it will shape the continuum to a redder slope at $\lambda<1.6{\rm~\mu m}$ but a bluer slope at $\lambda>1.6{\rm~\mu m}$, hence the kink shape. 

To describe this kink, we use a spectral index,
\begin{equation}
    m_{\rm kink}\equiv 0.498 m_{\rm J}-m_{\rm H}+0.502 m_{\rm K}\,, \label{eq:kink}
\end{equation}
where $m_{\rm J}, m_{\rm H}, m_{\rm K}$ are the magnitudes of the $\tilde{\rm J}$, $\tilde{\rm H}$, and $\tilde{\rm K}$ bands defined at the beginning of Section~\ref{sec:spectra}. The coefficients are chosen such that a high $m_{\rm kink}$ corresponds to a high $\tilde{\rm H}$-band flux relative to the two wings, while power-law SEDs of $F_\lambda\propto \lambda^0$ and $F_\lambda\propto\lambda^{-2}$ would give $m_{\rm kink}=0$.  Figure~\ref{fig:nearIR_triangle} shows the index values of our model spectra. At a given $T_{\rm eff}$, the kink strength increases with $\log g$. This is visually exemplified in Figure~\ref{fig:overview}, with the kink apparent for the two high-$\log g$ spectra but not seen for those with lower $\log g$. The gravity dependence is because a low atmosphere density tends to suppress the $\rm H^-$ population by Saha equilibrium \citep{Mihalas1978}. In addition, $m_{\rm kink}$ only weakly depends on metallicity at $T_{\rm eff}=5000{\rm~K}$. The metallicity will only introduce a scatter in $\log g$ by about one order of magnitude at a lower temperature of $T_{\rm eff}=4000{\rm~K}$, where the $\rm [M/H]=0$ curve shows a smaller index value than its lower-metallicity counterparts due to water absorption lines in the $\tilde{\rm H}$ band. We also check that $m_{\rm kink}$ is insensitive to $\xi_{\rm mtb}$, which introduces a scatter in $\log g$ by $<0.1$ dex at $T_{\rm eff}\geq4500{\rm~K}$ and $\log g<-1$. Overall, the kink can serve as a robust atmosphere density probe once $T_{\rm eff}$ is constrained from the overall SED. 

To our knowledge, the only LRDs with spectra up to $\lambda_{\rm rest}\sim2{\rm~\mu m}$ are the local ones reported in \citet{Lin2025b}. Long-wavelength spectroscopic coverage of high-redshift LRDs could test the optically thick atmosphere scenario on a demographic level, which we will discuss in Section~\ref{subsec:discussion_observation}.  Now, we compare the kink index of \egg\ with our models. Similarly to Section~\ref{subsec:near_IR_SED}, we present two measurements, first using the observed spectrum as-is, then removing the best-fit warm dust emission (Section~\ref{sec:application}) from the spectrum to represent the intrinsic emission of the gas atmosphere. The results are shown as the ``total'' and ``atmosphere'' lines in Figure~\ref{fig:nearIR_triangle}. 

The kink in \egg\ is remarkably weak, with $m_{\rm kink}=-0.062^{+0.005}_{-0.004}$ for the ``total'' measurement and $m_{\rm kink}=0.065^{+0.007}_{-0.006}$ for the ``atmosphere'' value.  As we have shown in Figures~\ref{fig:overview}--\ref{fig:color_color_IR} and will further illustrate in Section~\ref{sec:application}, we need $T_{\rm eff}\approx4500{\rm~K}$ to explain the optical-to-near-IR colors of \egg. Our model curve at this effective temperature only matches the intrinsic kink index at a very low $\log g\approx-3$. To further provide a stellar context, we measure the kink index of all stars in the second data release of the X-shooter Spectral Library \citep{Gonneau2020} with atmosphere parameters in the range of $4200{\rm~K}\leq T_{\rm eff}\leq5000{\rm~K}$ and $\log g\leq1.5$ \citep[evaluated and cataloged in][]{Arentsen2019}. These stars span a metallicity range of $-2.6<\rm[Fe/H]<0.1$ \citep{Arentsen2019}. All of the stars, with their relatively high gravity of $\log g>0$, show strong kinks of $m_{\rm kink}>0.2$, in agreement with the model points of similar parameters, and in contrast to \egg. We conclude that the lack of a kink feature in \egg\ implies $\log g\ll0$, i.e., a much lower photosphere density than those of giant stars. 

\subsection{The Balmer discontinuity}

\begin{figure}
    \centering
    \includegraphics[width=0.99\linewidth]{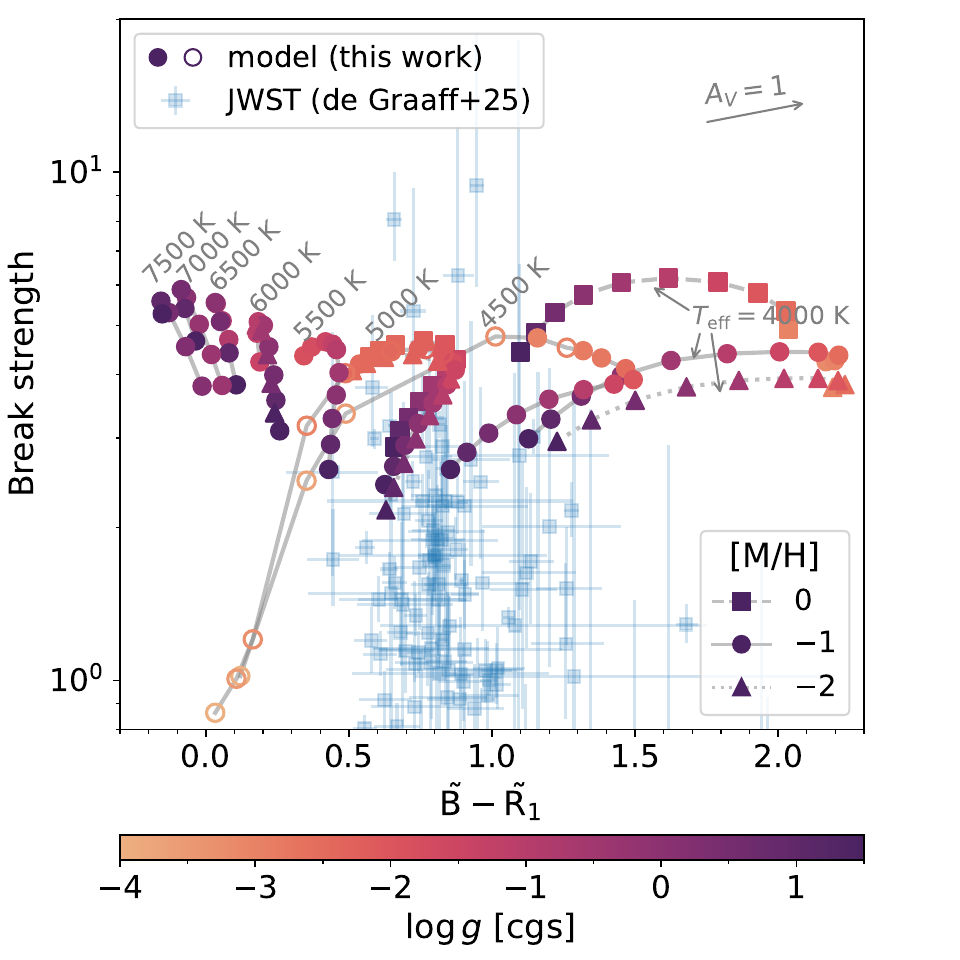}
    \caption{Balmer break strength and optical color of synthetic spectra and observed LRDs. Model points are color-coded by $\log g$ (which proxies the photospheric density); those with the same $T_{\rm eff}$ (annotated in gray) and [M/H] are connected by gray lines. Only models with $\xi_{\rm mtb}=2{\rm~km~s^{-1}}$ are shown here. The JWST LRD sample in \citet{deGraaff2025b} is shown in comparison. The model points will move down and approach the majority of the LRD data if there is a separate UV component contributing. The models are compatible with the break strengths of most observed samples but do not reproduce the strongest breaks.}
    \label{fig:break}
\end{figure}

The Balmer break, or a spectral flux discontinuity at $3646{\rm~\AA}$, is a prominent continuum feature in many LRDs and motivated the interpretation of the emission as being due to a gas layer surrounding the black hole \citep{Inayoshi2025,Ji2025,Naidu2025,Setton2024}. \citet{Liu2025} have shown that, as a result of the density dependence of the opacity across the Balmer limit, an atmosphere density much lower than that of typical stars can produce a Balmer break at a cool temperature of $T_{\rm eff}\sim5000{\rm~K}$, thus reconciling the break feature and the intrinsically red rest-optical color of LRDs.

The same qualitative trend holds in the spectra in this work. In Figure~\ref{fig:break}, for the model spectra, we measure their Balmer break strengths defined as in \citet{Liu2025}, i.e., the flux density ($f_\nu$) ratio between two bands, $[4000, 4100]\rm~\AA$ and $[3640,3645]\rm~\AA$. This definition follows that used in \citet{deGraaff2025,deGraaff2025b} but uses a narrower blue band in order not to traverse the theoretical Balmer limit. For models with $T_{\rm eff}\geq4500{\rm~K}$, at a given $T_{\rm eff}$, when $\log g$ decreases, the break strength first rises to a maximum before dropping. The maximum break strength is highest ($\sim6$) for $T_{\rm eff}=7000{\rm~K}$, whereas models with $T_{\rm eff}=4500{\rm~K}-5500{\rm~K}$ have maximum break strengths at $\sim4-5$. We note that this maximum is reached at a much lower $\log g$ for cool models. For example, the curve with $T_{\rm eff}=7500{\rm~K}$ peaks at $\log g=1.0$, or $\rho_{\rm ph}=1.2\times10^{-10}{\rm~g~cm^{-3}}$; but for $T_{\rm eff}=4500{\rm~K}$, the location of the maximum shifts to $\log g=-3.25$, or $\rho_{\rm ph}=4\times10^{-12}{\rm~g~cm^{-3}}$. This agrees with the finding in \citet{Liu2025} that cool atmospheres can give a Balmer break at low density. At $T_{\rm eff}=4000{\rm~K}$, the measured break strength becomes dominated by the $4000{\rm~\AA}$ break commonly seen in old galaxies rather than by the Balmer break itself. 

Observationally, the break strengths of JWST samples in \citet{deGraaff2025b} span a wide range. One possible interpretation invokes varying contributions from the host galaxy (which would not exhibit a prominent Balmer break if actively forming stars) and the atmosphere \citep{Barro2026,Sun2026}. Under such an interpretation, the models presented here are compatible with break strengths at or below $4-5$. However, the models fall below a few objects with the most extreme breaks \citep{deGraaff2025,Naidu2025}. The models also show weaker breaks than our previous calculations in \citet{Liu2025} at optical colors comparable to most LRDs. We note that the predicted break strength is sensitive to the atmosphere temperature profile. Numerically, this work calculates thermal equilibrium more rigorously than \citet{Liu2025}, where we used the gray atmosphere assumption and likely overestimated the break strength due to an overly cool upper atmosphere.  Therefore, we conclude that 1D plane-parallel atmosphere models are compatible with most but not all of the Balmer breaks of LRDs.  However, physically, this work only considers plane-parallel atmospheres, whereas a spherically extended atmosphere as in \citet{Liu2025} will naturally have cooler upper atmospheres due to geometric dilution and may have stronger breaks than in Figure~\ref{fig:break}.

Finally, in the bottom-left corner of Figure~\ref{fig:break}, our models with the lowest photosphere densities ($\rho_{\rm ph}<10^{-14}{\rm~g~cm^{-3}}$, or $\log g\sim-4$) show weak or even inverted breaks (Balmer jumps). This is physically possible in LTE models when the scattering opacity is important \citep{Schuster1905,Underhill1949}. Such a parameter space of low-density atmospheres may be relevant for objects such as the ``X-Ray Dot'' \citep{Hviding2026}, whose optical slope extends across the Balmer limit without signs of strong discontinuity. 

\subsection{Metal absorption lines}
\label{subsec:metal_absorption_lines}
\begin{figure}
    \centering
    \includegraphics[width=0.99\linewidth]{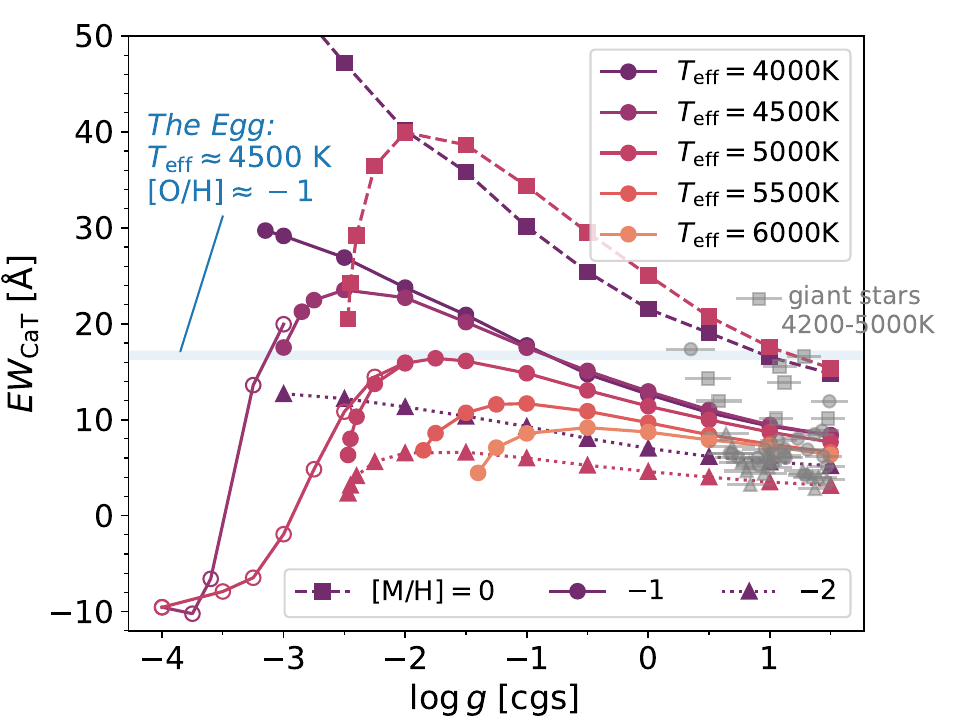}
    \caption{Total equivalent width of the CaT absorption lines ($\lambda\lambda8500, 8544, 8665\rm~\AA$). Three metallicities, $\rm [M/H]=0,-1,-2$, are shown for $T_{\rm eff}=4000, 5000{\rm~K}$, while only one metallicity, $\rm [M/H]=-1$, is shown for other values of $T_{\rm eff}$. Only $\xi_{\rm mtb}=2{\rm~km~s^{-1}}$ is shown, but $\xi_{\rm mtb}=10{\rm~km~s^{-1}}$ gives almost the same result because the EW is mainly contributed by the line wings. The blue horizontal band indicates the measurement of \egg\ \citep{Lin2025b}. Gray points show our measured CaT EW of the same giant star sample as in Figure~\ref{fig:nearIR_triangle}, but here we assume an $\alpha$-element enhancement of 0.4 dex for stars with $\rm[Fe/H]\leq-0.9$ and group them into metallicity bins of $\rm -0.5\leq[\alpha/H]<0.1$ (square), $\rm -1.5\leq[\alpha/H]<-0.5$ (circle), and $\rm -2.6<[\alpha/H]<-1.5$ (triangle) to compare to models. The CaT EW depends on the atmosphere density and thus on $\log g$, and the EW of \egg\ suggests $\log g\approx-3$ or $-1$. }
    \label{fig:ew_CaT}
\end{figure}

\begin{figure}
    \centering
    \includegraphics[width=0.99\linewidth]{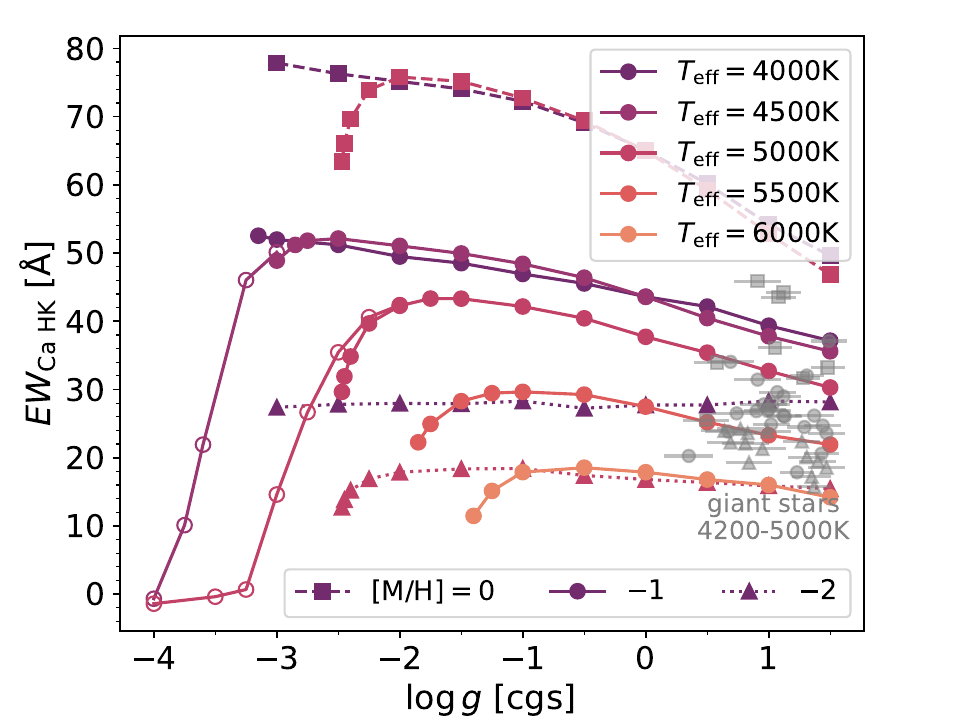}
    \caption{Same as Figure~\ref{fig:ew_CaT}, but for the total equivalent width of the Ca H and K lines ($\lambda\lambda3933,3968\rm~\AA$).}
    \label{fig:ew_CaHK}
\end{figure}

In our model spectra at $T_{\rm eff}\sim5000{\rm~K}$, the strongest metal absorption lines come from the \ion{Ca}{2} ion, including the Ca H and K lines ($\lambda\lambda3933,3968\rm~\AA$) in the optical and the Ca triplet (CaT, $\lambda\lambda8500, 8544, 8665\rm~\AA$) in the near-IR. These lines are well-known in stars at similar $T_{\rm eff}$ and have been identified in LRDs \citep{Juodzbalis2024,Ji2025b,Lin2025b}. Notably, the CaT equivalent width (EW) strongly anti-correlates with surface gravity (essentially the photospheric density) in the stellar regime and has been used as an atmosphere density probe \citep{Mallik1997,Cenarro2002,Mashonkina2007}.

We measure the total absorption EW of the CaT from our model spectra, as shown in Figure~\ref{fig:ew_CaT}. In the stellar regime, $\log g\gtrsim0$, we recover the established relation that the EW decreases with $\log g$ at a given $T_{\rm eff}$ and [M/H]. As pointed out in \citet{Jorgensen1992,Lin2025b}, with a decreasing atmosphere density, the continuum opacity is diminished more strongly than the CaT line opacity, leading to higher opacity contrast and thus higher EW. However, this trend reverses at very low $\log g$, and the CaT may even turn into emission ($EW_{\rm CaT}<0$). This is the regime where the scattering opacity becomes important, which can induce emission lines even in an optically thick medium in LTE \citep{Schuster1905,Underhill1949}. Because of this downturn, weak or absent CaT absorption does not necessarily indicate extremely low metallicity.

Figure~\ref{fig:ew_CaT} indicates significant challenges in using CaT alone to constrain $\log g$. The EW is sensitive to both $T_{\rm eff}$ and $\rm[M/H]$ besides $\log g$. Even if $T_{\rm eff}$ and $\rm[M/H]$ are known a priori, two $\log g$ values may give the same EW due to the non-monotonic dependence of $EW_{\rm CaT}$ on $\log g$. For example, \citet{Lin2025b} measured \egg\ to have $EW_{\rm CaT}=16.7\pm0.5{\rm~\AA}$, as indicated in Figure~\ref{fig:ew_CaT}; the gas-phase metallicity of the narrow line region has been independently measured as $\rm[O/H]=-1.33\pm0.08, -1.21\pm0.08$ \citep{Izotov2008}, $-1.11\pm0.03$  \citep{Lin2025b}, and $-0.96^{+0.21}_{-0.14}$ \citep{Ji2025b} (we scaled the original measurements to a solar abundance of $12+\log(\rm O/H)=8.69$ in \citealt{Asplund2009}). Assuming that the gas atmosphere of \egg\ has a similar metallicity of $\rm[M/H]=-1$, and taking $T_{\rm eff}=4500{\rm~K}$ from the SED shape of \egg, we find that either $\log g=-3$ or $\log g=-1$ gives an $EW_{\rm CaT}$ value consistent with data. However, we will argue in Section~\ref{sec:application} that we favor the $\log g=-3$ solution when combined with the continuum constraints. In either case, the $EW_{\rm CaT}$ value of \egg\ is higher than that of most giant stars at comparable effective temperatures and Ca abundances (gray circle points), which supports a low-density regime of $\log g<0$, as already argued in \citet{Lin2025b}.

We do a similar exercise for the total EW of the Ca HK lines, shown in Figure~\ref{fig:ew_CaHK}. The overall trend of the model spectra is similar to the CaT. Although the Ca HK lines have higher intrinsic EW than the CaT, the former are located at a shorter wavelength, where the host galaxy component and emission lines of LRDs likely bring significant infilling and complicate measurements in practice. We note that, without our implementation of partial frequency redistribution, the models would predict much broader HK lines, with the EW higher by 30\% for the case of $T_{\rm eff}=5000{\rm~K}, \log g=-2.0, {\rm[M/H]=-1}$ (Appendix~\ref{app:pfr}). Our treatment is still simplistic, and further development is needed \citep[see, e.g., ch.~13,][]{Mihalas1978} to study the detailed profile of such strong resonance lines, especially near the line core. 

\section{Application to a local LRD: the case for low black hole mass}
\label{sec:application}

\begin{figure*}
    \centering
        \includegraphics[width=0.99\linewidth]{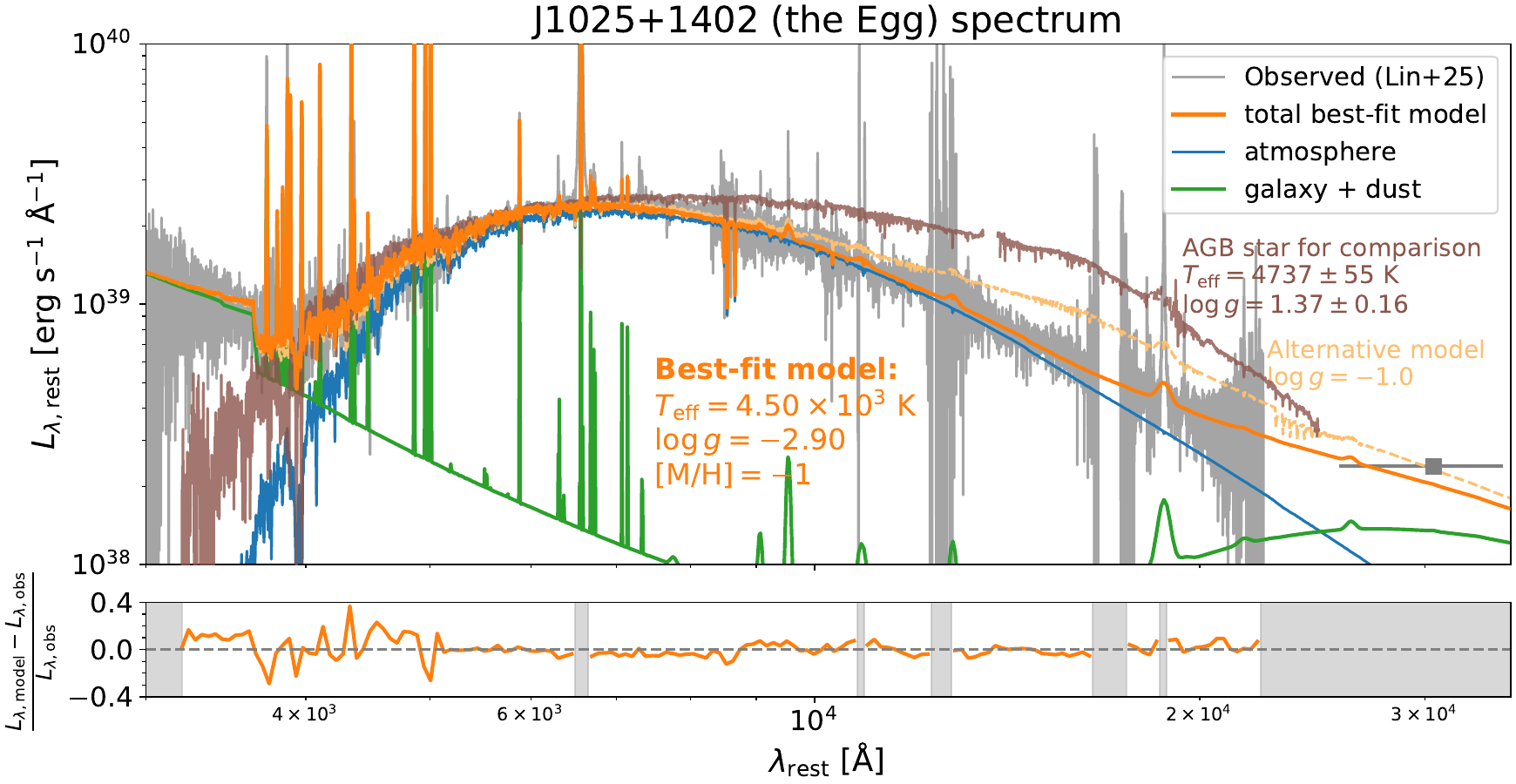}    
    \caption{Composite model for \egg\ compared to observation. Top: the model consists of an atmosphere at $T_{\rm eff}=4.50\times10^3{\rm~K}, \log g=-2.90, \rm[M/H]=-1$ reddened by an SMC dust law at $A_V=0.21$, a galaxy component, and a mid-infrared dust emission component. The low $\log g$ value is required by the narrow SED profile, the strong CaT absorption, and the lack of an $\rm H^-$ kink at $\lambda_{\rm rest}\sim1.6{\rm~\mu m}$.  In contrast, the low-resolution spectrum of an AGB star \citep[CL* NGC 6121 LEE 4302, $T_{\rm eff}=4737\pm55{\rm~K}, \log g=1.37\pm0.16, {\rm[M/H]}=-1.29\pm0.08$,][]{Arentsen2019} has a much broader SED and stronger $\rm H^-$ kink, which is also true for our alternative optically thick atmosphere model at $\log g=-1$. Bottom: residual of the best-fit model at low resolution normalized by the data luminosity. Shaded regions are excluded from the fitting due to strong emission lines or low data signal-to-noise ratio. }
    \label{fig:egg_spectrum_model}
\end{figure*}

\begin{figure}
    \centering
    \includegraphics[width=0.99\linewidth]{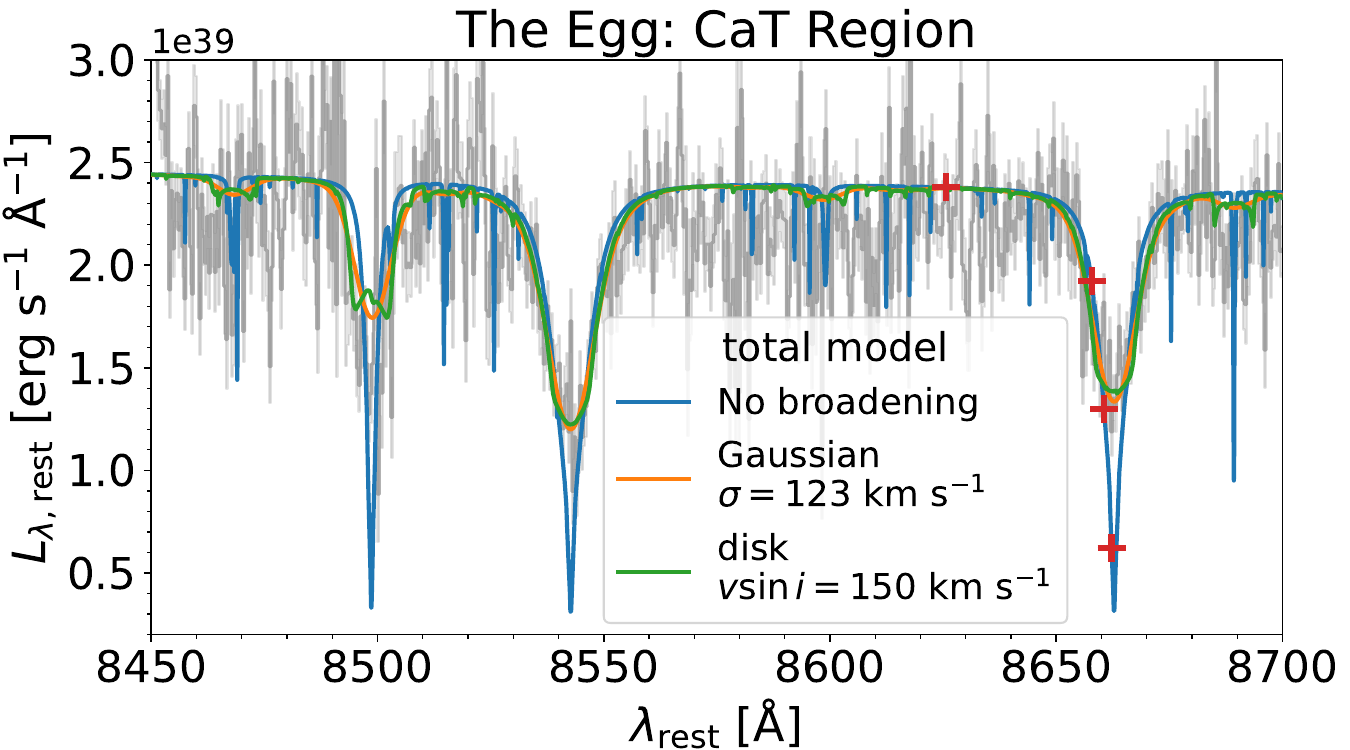}\\
    \vspace{2mm}
    \includegraphics[width=0.99\linewidth]{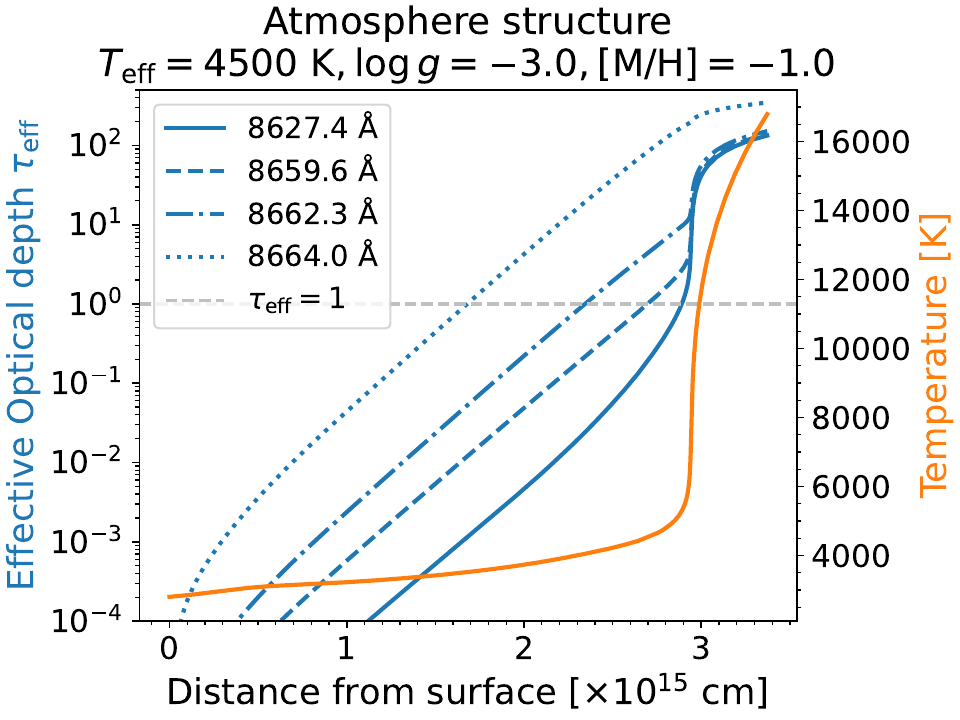}
    \caption{Top: same as Figure~\ref{fig:egg_spectrum_model}, but zoomed in to the CaT region. Three versions of the model spectrum are shown: without broadening, with a Gaussian broadening with $\sigma=123{\rm~km~s^{-1}}$, and with a disk-geometry broadening with $v\sin i=150{\rm~km~s^{-1}}$. The two broadening scenarios give subtly different line profiles. Red plus (``$+$'') symbols mark four wavelengths whose optical depths are shown in the bottom panel.  Bottom: Effective optical depth structure of a model atmosphere at $T_{\rm eff}=4500{\rm~K}, \log g=-3.0, {\rm[M/H]}=-1.0, \xi_{\rm mtb}=2\rm~km~s^{-1}$, whose parameters are close to the best fit to \egg. Blue curves show the optical depths at four wavelengths (solid: continuum; dashed, dot-dashed, and dotted: the unbroadened left wing of the rightmost CaT line, where the emergent flux is 3/4, 1/2, and 1/4 of the continuum flux). The legend gives vaccum wavelengths, which correspond to the red markers in the top panel (after a blueshift of $61{\rm~km~s^{-1}}$ to account for the observed CaT offset). The orange curve shows the temperature profile. The CaT lines and the adjacent continuum form at a radial separation much smaller than the inferred photosphere radius $R_{\rm ph}=8.7\times10^{15}{\rm~cm}$, which supports the assumption that the CaT line-forming region is approximately cospatial with the continuum photosphere. }
    \label{fig:egg_spectrum_CaT}
\end{figure}

\begin{figure*}
    \centering
    \includegraphics[width=0.99\linewidth]{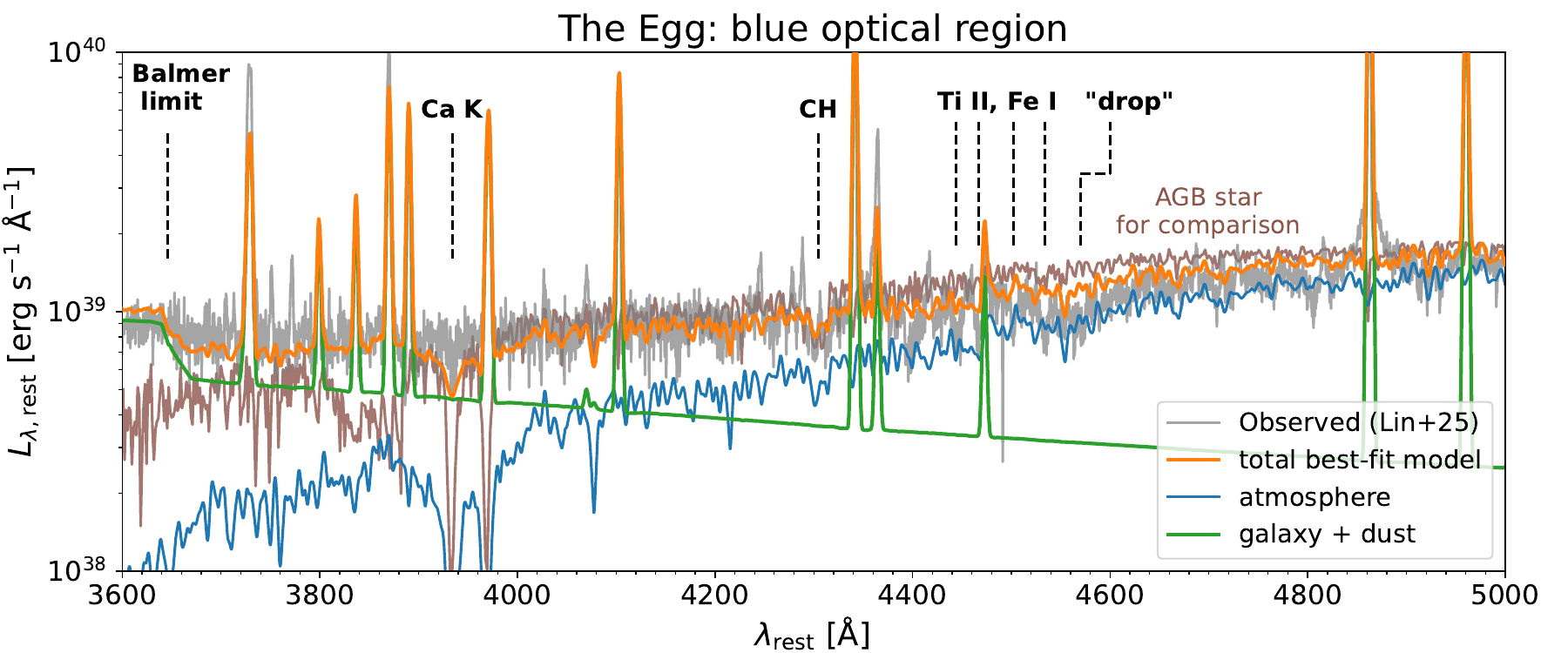}
    \caption{Same as Figure~\ref{fig:egg_spectrum_model}, but zoomed in to the blue optical region. }
    \label{fig:egg_spectrum_model_blue}
\end{figure*}

J1025+1402 \citep[``\egg'', $z_{\rm spec}=0.1007$,][]{Lin2025b,Ji2025b} is among the lowest-redshift LRDs known so far. In Section~\ref{sec:spectra}, we  discussed several aspects of its spectral features, all of which point to a low-density atmosphere. In this section, we focus on this object as a case study. We will present our best-fit model in Section~\ref{subsec:egg_bestfit}, where we bring together our previous analysis and highlight how a low $\log g$ simultaneously explains the narrow SED, the lack of a $\rm H^-$ kink, and the strong CaT absorption of \egg. Then, in Section~\ref{subsec:egg_interpretation}, we will interpret the low $\log g$ in terms of the atmosphere-scale total mass by outlining possible scenarios and discussing how to distinguish them in future modeling and observation. 

\subsection{Best-fit model and indications of low $\log g$}
\label{subsec:egg_bestfit}

We construct a composite spectral model of \egg. The total flux consists of three physical components:
\begin{enumerate}
    \item An atmosphere model with $T_{\rm eff}$, $\log g$, and $\xi_{\rm mtb}$ as free parameters. We fix $\rm[M/H]=-1$ for simplicity, which broadly agrees with the gas-phase metallicities measured previously \citep{Izotov2008,Lin2025b,Ji2025b}. 
    The atmosphere spectrum is convolved with a Gaussian kernel of $\sigma=123{\rm~km~s^{-1}}$ to match the CaT profile of the data. We pre-determined $\sigma$ by minimizing the $\chi^2$ difference at $8450{\rm~\AA}<\lambda_{\rm rest}<8700{\rm~\AA}$ (i.e., the CaT region) between the normalized observed spectrum and an atmosphere model at $T_{\rm eff}=4500{\rm~K},\log g=-3.0, {\rm[M/H]}=-1, \xi_{\rm mtb}=2{\rm~km~s^{-1}}$, which is close to our final best-fit result and has a CaT EW matching the data. Finally, we redden the spectrum by an SMC-averaged dust law \citep{Gordon2024} at a visual extinction magnitude of $A_V$.  The bolometric atmosphere luminosity before dust reddening is $L_{\rm atm}$. 
    \item A host galaxy, constructed using the code \textsc{FSPS} \citep{Conroy2009,Conroy2010,Byler2017}. We assume a \citet{Chabrier2003} initial mass function, a constant star formation rate\footnote{We also explored using a $\tau$-model for the star formation history and found that the best-fit $\tau$ far exceeded the stellar age.}, and a fixed metallicity of $\rm[M/H]_{\rm gal}=-1$. The model is parameterized by the stellar age $t_{\rm gal}$, the ionization parameter $\log U$ for the nebular emission, and the stellar mass $M_*$. We assume no dust extinction for the galaxy as implied by the very blue UV slope of \egg\ and the narrow Balmer decrement being fully consistent with the intrinsic Case B prediction \citep{Lin2025b,Ji2025b}; see also Appendix~\ref{app:sensitivity} where we test including dust extinction for the galaxy and find no significant change to our conclusion. 
    \item Warm dust emission, modeled as a blackbody with a luminosity $L_{\rm dust}$ at a fixed $T_{\rm dust}=1038{\rm~K}$. This approach follows \citet{Lin2025b} \citep[see also][]{Lyu2021}, who found that invoking a single blackbody temperature at $1038\rm~K$ reasonably described the SED in the optical-to-near-IR. Colder dust is only important at longer wavelengths and is neglected here.
\end{enumerate}

We have already argued in Section~\ref{sec:spectra} that only a low-$\log g$ model can simultaneously account for the narrow near-IR SED, the lack of the $\rm H^-$ kink, and the strong CaT absorption. To verify our expectation and to ensure that we have not missed potential degeneracies, we perform Markov-Chain Monte Carlo (MCMC) fitting to find the best-fit parameters, which we describe in Appendix~\ref{app:mcmc}. In the fitting only, we rebin the model and data to a low resolution of $\sim100$ to reduce noise correlation (Appendix~\ref{app:mcmc}); this implies that the fitting is mainly sensitive to the continuum. We emphasize that our MCMC posterior undoubtedly underestimates model systematics, so we refrain from interpreting the reported uncertainties. The purpose of our MCMC fitting is to identify the location and uniqueness of the best-fit solution with uninformed priors, which will support the robustness of a low $\log g$ for \egg. The prior on $\log g$ is restricted to the hydrostatic grid (Figure~\ref{fig:grid}, regime A of Section~\ref{subsec:super-Eddington_scenarios}), so the model is dynamically self-consistent at the photosphere, although our best fit requires dynamical modifications in the inner layers (see regime A of Section~\ref{subsec:super-Eddington_scenarios}). The no-$g_{\rm rad}$ models are excluded from the MCMC. In Appendix~\ref{app:sensitivity}, we further conduct sensitivity tests and demonstrate that the conclusion on low $\log g$ remains unchanged when assuming different metallicities, allowing $T_{\rm dust}$ to vary, applying independent dust extinction to the galaxy spectrum, or disregarding the WISE constraint. 

\begin{table}
    \centering
    \begin{tabular}{lcc}
        \toprule
        Parameter & Prior & Best-fit \vspace{1mm}\\
        \hline
        \textbf{Atmosphere} \\
        $T_{\rm eff}~(\rm K)$ & $[4000,5000]$ & $4500$ \\
        $\log (g/{\rm cm~s^{-2}})$ & hydrostatic & $-2.90$ \\
        $\xi_{\rm mtb}~(\rm km~s^{-1})$ & $[2,10]$ & $4$ \\
        $A_V$ & $[0,3]$ & $0.2$ \\        
        $\log (L_{\rm atm}/{\rm erg~s^{-1}})$ & $[41,45]$ & $43.35$ \\
        $\rm[M/H]_{\rm atm}$ & $-1$ (fixed) \vspace{1mm}\\
        \hline
        \textbf{Galaxy} \\
        $t_{\rm gal}~(\rm Myr)$ & $[1,12]$ & $1$ \\
        $\log U$ & $[-4,-1]$ & $-2.0$ \\
        $\log (M_*/M_\odot)$ & $[4,9]$ & $6.7$\\
        $\rm[M/H]_{\rm gal}$ & $-1$ (fixed) \vspace{1mm}\\
        \hline
        \textbf{Dust} \\
        $\log (L_{\rm dust}/{\rm erg~s^{-1}})$ & $[41,45]$ & $42.8$ \\
        $T_{\rm dust}~(\rm K)$  & 1038 (fixed) \vspace{1mm}\\
        \hline
        \textbf{Uncertainty} \\
        $\log s$ & $[-2,1]$ & $-1.1$ \\
        \bottomrule
    \end{tabular}
    \caption{Parameters in the composite model for \egg\ and their best-fit values from the MCMC run. Fixed parameters are distinguished from free ones by the ``fixed'' text in the prior. A ``hydrostatic'' prior for $\log g$ means the range is limited to the hydrostatic grid coverage in Figure~\ref{fig:grid}. The uncertainty parameter $\log s$ is defined in Appendix~\ref{app:mcmc}.}
    \label{tab:MCMC}
\end{table}

Table~\ref{tab:MCMC} summarizes the free parameters and their best-fit parameter values. Figure~\ref{fig:egg_spectrum_model} shows our composite model of \egg and the fractional fitting residuals. The displayed resolution of the residual is the same as what we use for the fitting. The green curve here gives the galaxy and dust components, which we have used in Section~\ref{sec:spectra}. First, we highlight the narrowness of the overall SED. In comparison, we plot the low-resolution spectrum of an Asymptotic Giant Branch (AGB) star, CL* NGC 6121 LEE 4302 \citep{Gonneau2020}, which has been inferred to have $T_{\rm eff}=4737\pm55{\rm~K}$ and $\log g=1.37\pm0.16$ \citep{Arentsen2019}. Its SED shows a slightly bluer optical slope than \egg\ but a dramatically redder color at $8000-16000{\rm~\AA}$.  In the X-shooter Spectral Library, the spectra of other giant stars at comparable $T_{\rm eff}$ appear similar. Based on the trend in Figure~\ref{fig:color_color}, we propose a much lower $\log g$ than typical of giant stars to solve the narrowness problem. We reiterate that dust reddening or microturbulence will not significantly influence the near-IR narrowness; on the other hand, radiative transfer effects at low atmosphere density naturally give the narrow SED.

Another important distinction between the SEDs of \egg\ and the AGB star is the strong $\rm H^-$ kink in the latter. In Figure~\ref{fig:egg_spectrum_model}, the slopes of the gray and brown curves are similar at $\sim2.0{\rm~\mu m}$, but the stellar SED visibly bends down at $\sim1.6{\rm~\mu m}$, while the SED of \egg\ continues without an obvious change in slope. This illustrates their difference in the kink index in Figure~\ref{fig:nearIR_triangle}. Again, a low value of $\log g=-3$ in our model suppresses the kink and gives a good agreement with data. This is the second piece of evidence for a low-density atmosphere.

Finally, we turn to the CaT absorption in \egg. As we showed in Figure~\ref{fig:ew_CaT}, the high CaT EW of this object, combined with $T_{\rm eff}\approx4500{\rm~K}$ and $\rm[M/H]\approx-1$, suggests two gravity solutions: $\log g=-3$ and $-1$. The orange curve in Figure~\ref{fig:egg_spectrum_CaT} (top panel), which is our best-fit model and includes the atmosphere at $\log g=-2.90$, illustrates the agreement of the model CaT EW with data\footnote{We blueshifted the model spectra by $61{\rm~km~s^{-1}}$, the measured CaT offset from the narrow emission lines in \egg\ \citep{Lin2025b}. In addition, the continuum flux in the data rises near the CaT region, and we are uncertain whether this is physical (e.g., from an extended atmosphere: see ch.~11-4, \citealt{Mihalas1978}) or results from imperfect calibration at the edge of the spectrum. To focus on the line profile, we scaled up the model curve by 15\% in the bottom panel to match the continuum of the data.}. We also explore the possibility of $\log g=-1$ by changing the atmosphere from $\log g=-2.90$ to $-1.00$ while keeping other parameters and components unchanged. The result is the yellow curve in Figure~\ref{fig:egg_spectrum_model}. It shows a significantly worse match at the near-IR, with the SED being too broad and the kink feature too strong. Therefore, although the CaT EW alone is subject to degeneracy between $\log g\approx-3$ and $-1$, the near-IR color and the absent H$^-$ kink unequivocally point to the low-density solution.

One may worry that inferring the intrinsic near-IR SED or the $\rm H^-$ kink strength of \egg\ requires modeling of the warm dust emission (Figures~\ref{fig:color_color_IR} and \ref{fig:nearIR_triangle}), which remains uncertain. However, the dust correction works in the opposite direction for the two probes. If dust emission were stronger than what we have found by MCMC, the intrinsic kink would be stronger and would suggest higher $\log g$, but meanwhile the $\rm \tilde{J}-\tilde{H}$ color in Figure~\ref{fig:color_color_IR} would become bluer and would suggest even lower $\log g$, and vice versa. We also directly demonstrate in Appendix~\ref{app:sensitivity} that our inference is robust to different assumptions of dust emission. We do caution that when modeling CaT absorption, we assumed that the atmosphere metallicity of \egg\ is similar to its gas-phase metallicity.  If the metallicity is larger (smaller) the inferred $\log g$ would decrease (increase).   Nevertheless, we stress that $\log g\approx-3$  can explain all the three spectral features discussed above. We note that the MCMC posterior touches the lower edge of $\log g$ in our model grid (Figure~\ref{fig:MCMC_corner}), which suggests that models at even lower $\log g$ beyond the grid coverage could potentially give similarly good fits. 

As a side note, since our model atmospheres do not emit strongly at short wavelengths, the UV and the blue optical emission (up to $\lambda_{\rm rest}\sim4100{\rm~\AA}$) in our best-fit model for \egg\ is dominated by the host galaxy, as shown in Figure~\ref{fig:egg_spectrum_model_blue}. Nebular emission from the host galaxy produces a small inverse Balmer break (higher flux at shorter wavelengths), which is broadly consistent with the data, despite the Balmer break in the atmosphere model. The deep Ca HK absorption in the atmosphere is also almost completely veiled, only leaving visible a weak line core of Ca K, although it still appears somewhat stronger than data. As we noted in Section~\ref{subsec:metal_absorption_lines}, our predictions near the Ca HK line cores are uncertain and require more detailed treatment. At longer wavelengths, the model reproduces the CH absorption feature at $\sim4305{\rm~\AA}$, an atmospheric feature as argued in \citet{Ji2025b}. The fitting is still imperfect: it underpredicts a series of \ion{Ti}{2} and \ion{Fe}{1} absorption lines, as well as the ``drop'' feature at $\sim4570{\rm~\AA}$ \citep{Ji2025b}. These features are also weak or absent from the spectrum of the AGB star. This suggests physical elements beyond the scope of this work, presumably gas absorption in the interstellar medium and/or turbulence in the atmosphere. We will further discuss the origin of the UV emission of LRDs in Section~\ref{subsec:discussion_observation}.

\subsection{Interpretation}
\label{subsec:egg_interpretation}
From Figure~\ref{fig:photosphere_density}, $T_{\rm eff}=4500{\rm~K}$ and $\log g=-3$ (close to the best fit in Table~\ref{tab:MCMC}) corresponds to $\rho_{\rm ph}=7\times10^{-12}{\rm~g~cm^{-3}}$. As we argued in Section~\ref{subsec:gravity}, the photosphere density is directly constrained by radiative transfer and is largely agnostic about dynamics. Indeed, the value of $\rho_{\rm ph}$ here agrees to within a factor of two with the one reported in \citet{Lin2025b} ($3.5\times10^{-12}{\rm~g~cm^{-3}}$), even though the earlier estimate did not assume hydrostatic equilibrium but instead was based on the super-Eddington accretion model in \citet{Liu2025}. In the following, we outline a few physical scenarios to step from $\rho_{\rm ph}$ to a central mass. We emphasize that the subsequent mass estimate is model-dependent and will discuss how the estimate changes when the invoked assumptions are varied. We note in passing that independent of $\rho_{\rm ph}$ or $\log g$, the photosphere radius (or the scale of the effective emitting surface in scenarios without a spherical geometry) is $R_{\rm ph}=\sqrt{L_{\rm atm}/4\pi\sigma T_{\rm eff}^4}=8.7\times10^{15}{\rm~cm}$, where we take $L_{\rm atm}=2.2\times10^{43}{\rm~erg~s^{-1}}$ and $T_{\rm eff}=4.50\times10^3{\rm~K}$ from Section~\ref{subsec:egg_bestfit}.

First, we disfavor the possibility that $g_{\rm dyn}=0$, i.e., that the atmosphere of \egg\ in fact satisfies radial hydrostatic equilibrium. Using $g_{\rm net}=g=10^{-3}{\rm~cm~s^{-2}}$ and applying Equation~(\ref{eq:gravity}), we obtain the sum of the black hole and gas mass enclosed by the photosphere as $M_{\rm tot}=6\times10^2~M_\odot$. However, assuming no radial density inversion, a gas sphere of the size of the photosphere has a minimum mass of $M_{\rm gas,min}=4\pi\rho_{\rm ph}R_{\rm ph}^3/3=9.6\times10^3~M_\odot>M_{\rm tot}$. We are then forced to invoke a gas shell whose radial thickness within the photosphere is limited to $\Delta R<2\times10^{14}{\rm~cm}\ll R_{\rm ph}$ by the total mass.  

The system can be less exotic if we relax the assumption of radial hydrostatic equilibrium and consider dynamical support due to bound motion of gas, i.e., $g_{\rm dyn}>0$ in Equation~(\ref{eq:hydro}). In this case, we interpret the fitting results as $g_{\rm net}=g-g_{\rm dyn}\sim10^{-3}{\rm~cm~s^{-2}}$. We refer to Equation~(\ref{eq:g_dyn}) and estimate the value of dynamical acceleration $g_{\rm dyn}$ by 
\begin{equation}
    g_{\rm dyn}={v_{\rm ph}^2\over R_{\rm ph}}=1.1\times10^{-2} {\rm~cm~s^{-2}}\,v_{\rm 100}^2\,, \label{eq:g_dyn_value}
\end{equation}
where $v_{\rm ph}$ is the characteristic velocity near the photosphere, and $v_{\rm 100}\equiv v_{\rm ph}/(100{\rm~km~s^{-1}})$. We see that $g_{\rm dyn}\gg g_{\rm net}$ for \egg\, if $v\gg30{\rm~km~s^{-1}}$. With this, we substitute $g \approx g_{\rm dyn}$ into the gravity term in Equation~(\ref{eq:gravity}) and obtain
\begin{equation}
    M_{\rm tot} = 6.5\times10^3~M_\odot \,v_{100}^2\,. \label{eq:mtot}
\end{equation}
This result does not depend on $g_{\rm net}$ itself other than the requirement of $g_{\rm dyn}\gg g_{\rm net}$. Therefore, the order of magnitude would remain correct even if $g_{\rm net}$ were larger by a factor of $\lesssim10$. 

The value of $v_{\rm ph}$ can be constrained using photospheric absorption lines depending on the nature of $v_{\rm ph}$. The gas kinematics broadly falls into three categories: radial ordered motion, i.e., advection in the form of contraction/expansion or infall/outflow; athimuthal ordered motion, i.e., rotation; and chaotic motion. The first case can be probed by line offsets; the second and third cases by line broadening. The absorption lines in \egg\ are offset from the narrow emission lines by $-100$ to $-60{\rm~km~s^{-1}}$ \citep{Lin2025b}, but it remains unclear whether this blueshift is due to gas motion at the atmosphere or to a peculiar velocity between the black hole and the narrow line region. The line widths suggest a velocity on the same order: in Figure~\ref{fig:egg_spectrum_CaT} (top panel), we find that assuming chaotic gas motion (with the spectrum convolved by a Gaussian kernel)\footnote{This is known as ``macroturbulence'', which pertains to gas motion over a scale comparable to or larger than the photosphere scale height and is distinct from microturbulence $\xi_{\rm mtb}$ \citep[e.g.,][]{Gray1978}. In practice, macroturbulence broadens the spectral flux and conserves line EWs, whereas microturbulence broadens the line opacity and in general changes the line EWs.}  gives $\sigma=123{\rm~km~s^{-1}}$, while assuming rotational broadening of a disk annulus gives a line-of-sight velocity of $v\sin i=150{\rm~km~s^{-1}}$, where $i$ is the inclination angle of the disk. Our $\sigma$ is less than what one may naively infer from the FWHM \citep[$441\pm18{\rm~km~s^{-1}}$,][]{Lin2025b} because of the already large intrinsic line width (blue curve in Figure~\ref{fig:egg_spectrum_CaT}, top panel) required to explain the EW. 

If the atmosphere is spherical, then we expect $v_{\rm ph}$ to be on the same order as the measured broadening. With $v_{\rm ph}\sim1.2\times10^2{\rm~km~s^{-1}}$, Equation~(\ref{eq:mtot}) implies that $M_{\rm tot}\sim1\times10^4~M_\odot$, on the same order as the gas mass $M_{\rm gas,min}$. Therefore, this scenario features a small black hole mass, $M_{\rm BH}\leq1\times10^4~M_\odot$, and the gravity of the system may be dominated by the gas sphere instead of the black hole itself.  Notably, the Eddington ratio of this system will be $\lambda_{\rm Edd}=L_{\rm atm}/L_{\rm Edd}(M_{\rm BH})\geq2\times10^1$. We also note that the dynamical timescale implied by the mass, $t_{\rm dyn}\equiv \sqrt{R_{\rm ph}^3/GM_{\rm tot}}=3\times10^1{\rm~yr}\, v_{\rm 100}^{-1}$, is longer than the rest-frame duration over which no strong variability is detected for \egg\ \citep[$\sim5$ yr in the $r$ band,][]{Burke2025}. 

If, on the other hand, the atmosphere is in a disk geometry, then $v_{\rm ph}$ may be greater than the measured line-broadening velocity by a factor of $(\sin i)^{-1}$. This may allow for a much larger total mass. In addition, the minimum gas mass of the disk will likely be less than our previous estimate by a factor of the disk aspect ratio, leaving the black hole likely dominating the mass budget. Therefore, this disk scenario will have $M_{\rm BH}\sim1\times10^4(\sin i)^{-2}~M_\odot$, with the Eddington ratio $\lambda_{\rm Edd}\sim2\times10^1(\sin i)^2$. Even so, we expect a lower limit on $\sin i$ at the disk aspect ratio $H/R\sim0.1$ for a geometrically thin, perfectly face-on disk, which implies $M_{\rm BH}\leq1\times10^6~M_\odot$. The higher $v_{\rm ph}$ in this scenario would imply a shorter dynamical timescale, likely with an even shorter thermal timescale for a gravitationally unstable disk \citep{Thompson2005}, but whether variability at such timescales is observable depends on the number of embedded sources that power the disk emission, which is uncertain. 

The sphere and disk scenarios may be observationally distinguishable from the line profiles, as shown in Figure~\ref{fig:egg_spectrum_CaT} (top panel) for the CaT. Disk rotation tends to produce double-peaked or boxy line profiles. This effect is most easily seen in relatively weak lines, e.g., the leftmost CaT line at $\lambda_{\rm rest}=8500{\rm~\AA}$. Future observation with a higher signal-to-noise ratio and spectral resolution may probe the detailed line shapes and directly constrain the atmosphere geometry. However, we caution that a lack of a double-peaked profile does not necessarily rule out the disk scenario if the disk is not geometrically thin or has optically thick winds \citep[cf. the formation of single-peaked emission lines from disk winds in][]{Murray1997}.

Now, we discuss the uncertainties in the above-quoted constraint on $g_{\rm dyn}$ and hence $M_{\rm tot}$. First, the observed CaT broadening is assumed to reflect gas motion rather than electron scattering \citep[e.g.,][]{Munch1948} or other broadening mechanisms. If the latter are important, the actual $M_{\rm tot}$ could be even lower than inferred above. However, our mass contraint also assumes that the CaT line-forming region is cospatial with the continuum photosphere. The latter is consistent with our model setup, where the atmosphere scale height is small relative to the photosphere radius (Equation~(\ref{eq:h_ph})). To further test co-spatiality, we select a few wavelength points on the left wing of the Ca II $\lambda8665{\rm~\AA}$ line and the adjacent continuum in the best-fit Egg model in Figure~\ref{fig:egg_spectrum_CaT}, top panel, and display their effective optical depth ($\tau_{\rm eff}$) profiles in the bottom panel. The location where $\tau_{\rm eff}=1$ represents where the spectrum ``forms''; in this sense, the line wing forms at a distance $\lesssim1\times10^{15}{\rm~cm}$ from the continuum-forming location. This is much less than the photosphere radius $R_{\rm ph}=8.7\times10^{15}{\rm~cm}$ inferred above, supporting co-spatiality within our model framework. On the other hand, if the real atmosphere density profile deviates from our model and is spherically extended, with the line forming region located at a radius $R_{\rm CaT}>R_{\rm ph}$, the estimates in Equations~(\ref{eq:g_dyn_value})(\ref{eq:mtot}) will be greater by a factor of $R_{\rm CaT}/R_{\rm ph}$. Still, CaT absorption requires dense gas at a relatively high temperature. The metastable 3d level of Ca II that provides the absorption needs to be collisionally populated by an electron density $n_e\gg n_{\rm crit}\sim10^7{\rm~cm^{-3}}$ (where the critical density is estimated from an Einstein A value of $0.9{\rm~s^{-1}}$ of the 3d $\to$ 4s transition and a collision strength of $4\times10^0$; data from the CHIANTI database, \citealt{Dufresne2024}, originally from \citealt{Melendez2007}), which is not far from the LTE electron density at the photosphere ($n_e\sim10^9{\rm~cm^{-3}}$ for gas at $T=4500{\rm~K}$ and $\rho=10^{-11}{\rm~g~cm^{-3}}$). Additionally, gas significantly colder than $4500$~K cannot efficiently produce the absorption, as the CaT line opacity at LTE would decrease by a factor of 10 when $T=3000$~K. We thus deem it plausible that the CaT line is formed close to the continuum photosphere even if the atmosphere is extended, while acknowledging that the actual $M_{\rm tot}$ could be higher than the limit we quoted if the CaT forms much further out. 

\section{Summary and discussion}
\label{sec:discussion}
Understanding the nature of LRDs requires bridging the gap between detailed spectroscopic observations and theories attributing their optical-to-near-IR emission to an optically thick atmosphere. The commonly adopted blackbody approximation, while useful, misses spectral details that are essential for testing and applying these theories. To address this limitation, we construct a synthetic spectral library of optically thick atmospheres over a parameter space tailored to LRDs, which we have made publicly available on GitHub (Section~\ref{sec:intro}). We systematically explore the dependence of key continuum and line features on atmospheric parameters, most importantly the surface gravity (a proxy for the photospheric density; see Figure~\ref{fig:photosphere_density} and Section~\ref{subsec:gravity}); then, applying the library to a local LRD, \egg, we infer a low photospheric density $\rho_{\rm ph} \sim 10^{-11} {\rm~g~cm^{-3}}$ (or $\log g\approx-3$), which, under additional dynamical assumptions, suggests a small total mass within the photosphere. This work provides a foundation for further tests of the optically thick atmosphere scenario and suggests a way of constraining the black hole mass within this picture. Our main conclusions are as follows.

\begin{enumerate}
    \item The SEDs of optically thick atmospheres deviate from blackbodies. The optical and near-IR SEDs may appear broader or narrower than a blackbody, and the color temperatures may differ from $T_{\rm eff}$ by $>1000{\rm~K}$ (Figures~\ref{fig:color_color} and \ref{fig:color_color_IR}), depending on atmospheric parameters. Theoretical modeling of LRDs needs to consider radiative transfer effects instead of assuming simple blackbody emission.
    \item The near-IR continuum provides useful constraints on photospheric density, i.e., on $\log g$. The $\rm \tilde{J}-\tilde{H}$ color and the $\rm H^-$ kink strength increase monotonically with $\log g$, avoid the degeneracy from microturbulence, and are measurable from low-resolution spectroscopy (Figures~\ref{fig:color_color_IR} and \ref{fig:nearIR_triangle}; see also Figure~\ref{fig:egg_spectrum_model}). 
    \item Cool atmospheres ($4500\leq T_{\rm eff}\leq6000{\rm~K}$) can produce a Balmer break at $\log g\ll0$, with a peak break strength of $4-5$ (Figure~\ref{fig:break}). The peak strength is consistent with most LRDs but falls short of those with the strongest breaks.
    \item At a given $T_{\rm eff}$ and $\rm[M/H]$, the EWs of the CaT and Ca HK absorption lines first increase and then decrease as the photospheric density ($\log g$) becomes lower (Figures~\ref{fig:ew_CaT} and \ref{fig:ew_CaHK}). Synthetic spectra at moderately low $\log g$ may give stronger EWs compared to the stellar regime ($\log g \gtrsim0$), but those at very low $\log g$ show weak absorption or even emission. 
    \item The observed spectrum of \egg\ appears narrower than a blackbody in the optical to near-IR (Figures~\ref{fig:color_color} and \ref{fig:color_color_IR}), has a weak $\rm H^-$ kink (Figure~\ref{fig:nearIR_triangle}), and shows strong CaT absorption for its metallicity (Figure~\ref{fig:ew_CaT}). These are simultaneously explained by a low value of $\log g \approx-3$ at $T_{\rm eff}=4.50\times10^3{\rm~K}$, which yields a photospheric density of $\rho_{\rm ph}\approx 10^{-11}{\rm~g~cm^{-3}}$ (Figure~\ref{fig:photosphere_density}). This is corroborated by MCMC fitting to a composite model (Figure~\ref{fig:egg_spectrum_model}, Table~\ref{tab:MCMC}). The CaT EW alone admits two solutions ($\log g\approx-3$ and $\log g\approx-1$), but the narrow near-IR SED and the absent H$^-$ kink unequivocally favor the low-density solution (Section~\ref{subsec:egg_bestfit}). 
    \item The low photospheric density ($\rho_{\rm ph}\approx10^{-11}{\rm~g~cm^{-3}}$) is a direct result of the radiative transfer modeling and disfavors radially hydrostatic configurations. Translating this into a central mass is more model-dependent. Under the additional assumption that the CaT line width traces turbulent gas velocity and that the CaT line-forming region is cospatial with the continuum-forming layer, we estimate $g_{\rm dyn}$ from the CaT broadening (Equations~(\ref{eq:g_dyn_value})(\ref{eq:mtot})) and infer $M_{\rm BH}\lesssim1\times10^4~M_\odot$ in a spherical geometry or $M_{\rm BH}\lesssim1\times10^6~M_\odot$ in a face-on disk geometry, corresponding to Eddington ratios $\lambda_{\rm Edd}=L_{\rm atm}/L_{\rm Edd}\geq20$ and $\lambda_{\rm Edd}\geq0.2$ respectively (Section~\ref{subsec:egg_interpretation}). The mass limit will be higher if the CaT forms exterior to the continuum photosphere.
\end{enumerate}

\subsection{Caveats and future directions for modeling}
\label{subsec:discussion_model}
Our spectral library does not explain all features of LRDs and does not necessarily apply to all (or any!) LRDs. Our calculations are designed for atmospheres with sufficiently high optical depth, whose deep layers thermalize any radiation and keep atomic level populations at LTE.\footnote{This is similar to what is termed the ``blackbody envelope'' scenario of LRDs in \citet{Asada2026}, although we refer to our model the ``optically thick atmosphere'' because (a) we have shown that the spectrum of an optically thick atmosphere has features deviating from a blackbody, and (b) our model is not limited to a spherical geometry.} The inference of $\log g$ is based on the optically thick assumption. Such atmospheres in our current framework do not explain the broad Balmer emission lines. Our synthetic spectra also have difficulty reproducing the strongest observed Balmer breaks of LRDs. However, one important feature of optically thick atmospheres is that they robustly predict many spectral features and correlations between them that can be used to test the model (just as in stars).   In addition, 
the optically thick atmosphere model naturally produces narrow continuum SEDs, as is strikingly seen in \egg. 

The optically thick atmosphere can be contrasted with the ``cocoon envelope'' scenario \citep{Naidu2025,Rusakov2026,Asada2026,Sneppen2026}, which features a gas layer that partially reprocesses ionizing incident radiation and produces broad emission lines in situ. In this case, the radiation and level populations throughout the gas layer can be markedly different from LTE and are not described by the models in this paper. However, the optically thick atmosphere and the cocoon are not mutually exclusive. Our models have higher photosphere volume and column densities ($n_{\rm H}\sim10^{12}{\rm~cm^{-3}}$ and $N_{\rm H}\sim10^{27}{\rm~cm^{-2}}$ in the best-fit model for \egg) than typical cocoon models \citep[e.g., $n_{\rm H}\sim10^{10}{\rm~cm^{-3}}$ and $N_{\rm H}\sim10^{24-26}{\rm~cm^{-2}}$ in][]{Sneppen2026}, but our fitting does not rule out a low $n_{\rm H}$ since we have not fully covered that part of the parameter space (e.g., $\log g\ll-3$ at $T_{\rm eff}<4500{\rm~K}$) due to convergence issues. Furthermore, the cocoon model spectra in \citet{Sneppen2026} with the highest column densities have already become optically thick in the Balmer continuum (evidenced by their strong Balmer breaks) and moderately so in the optical (evidenced by their continuum becoming reminiscent of a blackbody and their Ca HK absorption). The cocoon picture at still higher values of $n_{\rm H}$ and/or $N_{\rm H}$ will likely smoothly connect to our LTE optically thick atmosphere regime. It would be very valuable to quantitatively explore this transition in order to test both models against LRD observations.    In addition, while the cocoon models reproduce the broad emission lines, the spectra in \citet{Sneppen2026} still require substantial dust extinction ($A_V>1$ mag; their Figure B8) to reproduce the observed variety of SED curvature in the optical, especially at the narrow end. We speculate that a wide range of volume and/or column densities spanning the optically thick atmosphere regime and the cocoon regime may simultaneously show broad emission lines and a narrow SED. Future demographic studies in the rest-near-IR may further test which regime is more applicable to the continuum, which we will discuss in Section~\ref{subsec:discussion_observation}. 

The optically thick atmosphere is inherent in various physical pictures of LRDs, such as quasi-stars \citep{Begelman2025,Santarelli2025}, spherical super-Eddington accretion flows \citep{Liu2025}, supermassive stars \citep{Nandal2026}, and gravitationally unstable disks \citep{ZhangC2025,Zwick2025,Chen2026}. For example, \citet{Santarelli2025} predicted $\log g<0$ from their quasi-star models, but they used stellar synthetic spectra at $\log g=0$ because lower $\log g$ were then unavailable. Our spectral library, which covers the parameter space of their model LRDs, can be directly incorporated into their framework and indeed into future modeling works. Specifically, our library requires particular photospheric temperature, density and metallicity to match data, which may or may not be plausible in the various proposed scenarios. For example, the quasi-star radius (and hence atmosphere properties) is set by saturated convection, whereas the supermassive star scenario may have distinct predictions depending on its evolutionary and pulsational properties; these models may also expect different elemental abundance patterns \citep{Begelman2025,Nandal2026b}. We caution that our library alone cannot tell a gas-enshrouded black hole from a supermassive star if they predict identical atmospheres. The library would be most effective when combined with a dynamical model (as we simplistically did for \egg), along with constraints beyond the spectrum, such as variability, occurrence rate and clustering. 

Within the regime of the optically thick atmosphere, we have conducted (admittedly limited) tests on non-LTE effects and found only mild deviations from the LTE synthetic spectra. We calculate non-LTE level populations for H, Mg, and Ca in two cases, $\log g=0$ and $-2$ (with $T_{\rm eff}=5000{\rm~K}, {\rm[M/H]=-1}, \xi_{\rm mtb}=2{\rm~km~s^{-1}}$). Compared to LTE, all photometric band magnitudes differ by less than 0.03 mag, and the Balmer break strength and the CaT EW are enhanced by $<1.5\%$ and $<12\%$ in both cases. This agrees with previous calculations showing weak non-LTE effects on the CaT in the stellar regime \citep{Jorgensen1992,Mashonkina2007,Merle2011}. We caution, however, that extrapolating these to $\log g\approx-3$ (near the best-fit model for \egg) introduces additional uncertainty that we have not quantified. In addition, our tests are restricted in allowing only a few species to have non-LTE populations and in fixing the density and temperature to the LTE profile. In particular, we still assume an LTE number density for $\rm H^-$. This assumption is relevant for the H$^-$ kink diagnostic, which is one of the primary indicators of low density in \egg: if the number density of H$^-$ significantly deviates from LTE, the predicted kink strength could differ from our calculation. Despite this uncertainty,  \citet{Barklem2024} recently evaluated non-LTE effects of $\rm H^-$ for stars and found a $<1\%$ difference in the optical continuum for $T_{\rm eff}\leq5000{\rm~K}$ and $\log g\geq1$, with a tendency to overpopulate $\rm H^-$ compared to LTE. If this tendency holds at lower $\log g$, we will expect the $\rm H^-$ opacity to be more manifest in non-LTE (e.g., stronger kinks), which will strengthen our conclusion about the low $\log g$ of \egg.  An even greater uncertainty is that the predicted spectra may be modified by additional physics, e.g., photoionization by massive stars in the vicinity or shock heating due to accretion, which have been suggested as potential origins of emission lines for LRDs \citep{Baggen2024,Asada2026,Chen2026}. 

We have made several simplifications about the gas distribution in our atmosphere models. The 1D model neglects spatial inhomogeneity and requires simplified treatments such as mixing-length convection and microturbulence. The plane-parallel geometry will break down for extended atmospheres. The dynamical implications of gas motion are only considered in an order-of-magnitude fashion. Multi-dimensional radiation hydrodynamic simulations will be needed to capture these complications self-consistently, and our calculations provide a baseline for comparison with future, more dynamically realistic models. In the alternative cocoon scenario, simulations could clarify the formation mechanism and stability of the gas cocoon and connect it to the accretion process of the black hole. 

\subsection{Future directions for observation}
\label{subsec:discussion_observation}

Throughout this work, we have chosen \egg\ as a key observational point of comparison. The near-IR spectral coverage up to $\lambda_{\rm rest}\sim2{\rm~\mu m}$ is crucial in our inference of $\log g$ (via the $\rm\tilde{J}-\tilde{H}$ color and the $\rm H^-$ kink strength). In addition, the high-resolution, high signal-to-noise ratio spectrum of metal absorption lines, in particular the CaT, provides additional evidence for a low $\log g$ and enables us to constrain the dynamical acceleration $g_{\rm dyn}$ via line broadening.   The resulting atmosphere-scale total mass we find for \egg\ is $\sim 10^{4-6}~M_\odot$, with the lower value applying in a spherical model and the higher value applying for a face-on disk.   The black hole mass can be two orders of magnitude lower than that inferred from virial scaling relations \citep[$10^{6.5}~M_\odot$,][]{Lin2025b}. A similar deviation has been suggested for many high-redshift LRDs based on detailed modeling of the broad emission lines \citep{Rusakov2026,Naidu2025,Chang2026,Torralba2025,Sneppen2026}, although debates remain \citep{Juodzbalis2024,Brazzini2025,Brazzini2026}. Future analysis of the broad emission lines of \egg\ may provide independent constraints on its central engine. Our estimate for \egg\ suggests a rapidly accreting black hole in the intermediate mass regime, which, if further confirmed, will have profound implications for the demographics and evolutionary history of massive black holes.

The methodology here in principle can be applied to JWST LRDs at higher redshifts under the assumption that the optical-to-near-IR emission arises from an optically thick atmosphere, albeit with large uncertainties due to the currently limited spectroscopic wavelength coverage and the scarcity of high-frequency-resolution data. 
In the case of \textit{the Rosetta Stone}, we have inferred $\log g\leq-1.0$ in Section~\ref{subsec:near_IR_SED}; this  implies $M_{\rm tot}\leq8\times10^5~M_\odot\,(L_{\rm atm}/3\times10^{44}{\rm~erg~s^{-1}})(T_{\rm eff}/4500{\rm~K})^{-4}$ assuming radial hydrostatic equilibrium (Equation~(\ref{eq:gravity})), but we caution that (a) here we have no strong constraints on $g_{\rm dyn}$, which is important in the case of \egg; and (b) the reference value of the atmosphere luminosity $3\times10^{44}{\rm~erg~s^{-1}}$ is measured directly from the spectrum at $3600{\rm~\AA}<\lambda_{\rm rest}<16700{\rm~\AA}$ without correcting for dust extinction or subtracting the UV or dust emission components. 

\begin{figure}
    \centering
    \includegraphics[width=0.95\linewidth]{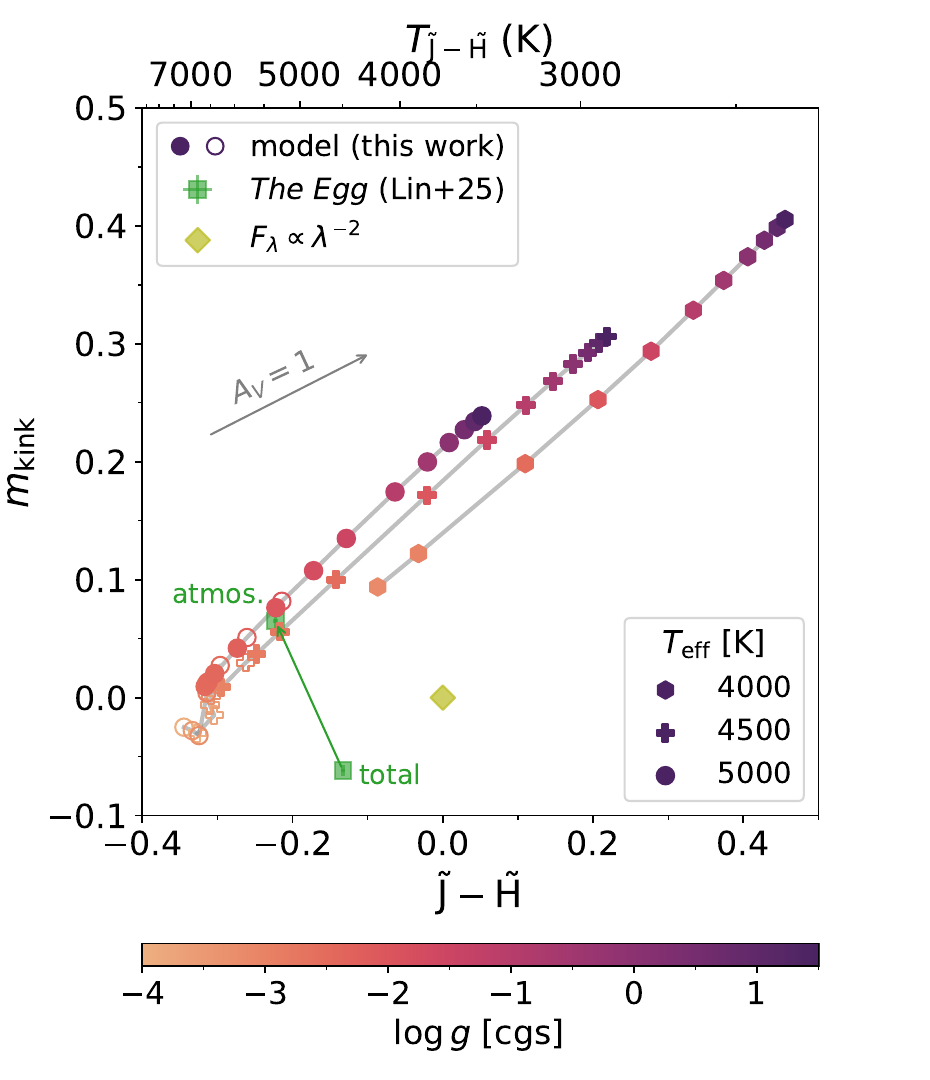}
    \caption{Relation between the near-IR color and the $\rm H^-$ kink strength, similar to Figure~\ref{fig:color_color_IR}. Only models with $\rm[M/H]=-1$ and $\xi_{\rm mtb}=2{\rm~km~s^{-1}}$ are shown. The measurement for \egg\ is shown in comparison, where the ``total'' point measures the original spectrum and the ``atmos.'' point measures the spectrum subtracted by a best-fit young galaxy model dominating in the UV and a warm dust emission model dominating in the mid-IR (Section~\ref{sec:application}). Gray vector indicates the change by an SMC dust extinction of $A_V=1$. The optically thick atmosphere scenario predicts a correlation between $\rm\tilde{J}-\tilde{H}$ and $m_{\rm kink}$ for the LRD population, which may be distinguished from a power-law SED $F_\lambda\propto\lambda^{-2}$. }
    \label{fig:color_color_IR_kink}
\end{figure}

Even before attempting to infer the black hole mass for individual JWST samples, one may first want to test whether the optically thick assumption holds on the population level. A prediction of the optically thick atmosphere scenario is a population-wise correlation between $\rm\tilde{J}-\tilde{H}$ and $m_{\rm kink}$ since both increase with the $\log g$ parameter, as shown in Figure~\ref{fig:color_color_IR_kink}. Furthermore, presuming that $g_{\rm net}$ correlates with $M_{\rm BH}$ and that $M_{\rm BH}$ correlates with bolometric luminosity, we expect more luminous LRDs to appear broader in the near-IR SED, have redder $\rm\tilde{J}-\tilde{H}$, and show stronger kinks. This trend can be distinguished from the prediction currently presented in \citet{Sneppen2026}, which generally shows a power-law near-IR SED with a typical slope of $F_\lambda\propto \lambda^{-2}$ (diamond symbol in Figure~\ref{fig:color_color_IR_kink}),\footnote{\citet{Sneppen2026} noted that they currently did not include the $\rm H^-$ opacity in their calculation, so the predictions are not strictly comparable. However, the lower typical gas density in their model likely makes $\rm H^-$ less important than in our models. } although we caution that dust emission may weaken the observed kink strength (as in our model for \egg). 

Future observations could allow more conclusive modeling for high-redshift LRDs. First, a JWST/MIRI spectroscopic survey could cover the near-IR continuum including the kink feature at $z>2$, where most known LRDs are located. The near-IR continuum could test the optically thick scenario (Figure~\ref{fig:color_color_IR_kink}). If confirmed, joint modeling of dust and atmosphere emission can be used to measure $\log g_{\rm net}$ as we have done for \egg. Second, high-resolution spectroscopy aimed at potential absorption lines, e.g., the CaT, could simultaneously constrain $g_{\rm dyn}$ from line broadening and $g_{\rm net}$ from line EWs. A combination of these two observations could give a black hole mass measurement independent of the one from emission lines. Our work also motivates the search for more low-redshift LRDs, whose rest-near-IR spectra are accessible with JWST/NIRSpec or even by ground-based IR telescopes, as is the case in \citet{Lin2025b}. A larger sample size could address whether the atmospheric condition of \egg\ is common or special. 

The optically thick atmosphere models in this work do not emit strongly in the UV, which necessitated the galaxy component in our fitting of \egg\ to dominate the UV and blue optical. This is compatible with the spatially resolved UV emission of some LRDs \citep{Rinaldi2024,Chen2025a,Chen2025b} and the scarcity of evidence for short-term UV variability \citep{Kokubo2024,ZhangZ2025}, and similar scenarios have been considered in some demographic studies of LRDs \citep[e.g.,][]{Barro2026,Sun2026}. However, it remains possible that the central engine contributes to the UV, as in the cocoon scenario at relatively low column density \citep{Sneppen2026}. In addition, a nearby blue companion could supply part of the UV flux in a subset of LRDs \citep{Baggen2025,Baggen2026}. Future image decomposition analysis and variability monitoring of LRDs will bring a more comprehensive understanding of the entire SED.

\begin{acknowledgments}
We thank the anonymous referee for their comments and suggestions, which substantially improved this paper. HL thanks Ivan Hubeny for providing the water line list used in this work and for technical help. HL thanks Rohan Naidu, Anna de Graaff, David Setton, and Bingjie Wang for constructive discussions. The calculations presented in this article were performed on computational resources managed and supported by Princeton Research Computing, a consortium of groups including the Princeton Institute for Computational Science and Engineering (PICSciE) and the Office of Information Technology's High Performance Computing Center and Visualization Laboratory at Princeton University. The Center for Computational Astrophysics at the Flatiron Institute is supported by the Simons Foundation.
\end{acknowledgments}

\software{Astropy \citep{astropy2013,astropy2018,astropy2022}, Claude Code \citep{Anthropic2025}, corner.py \citep{Foreman-Mackey2016}, emcee \citep{Foreman-Mackey2013}, FSPS \citep{Conroy2009,Conroy2010,Byler2017}, Matplotlib \citep{Hunter2007}, Numpy \citep{Harris2020}, Pyphot \citep{Fouesneau2025}, Scipy \citep{Scipy2020}, TLUSTY \citep{Hubeny1988,Hubeny2021}}

\appendix
\section{Partial frequency redistribution of Ca H and K lines}
\label{app:pfr}
\begin{figure}
    \centering
    \includegraphics[width=0.99\linewidth]{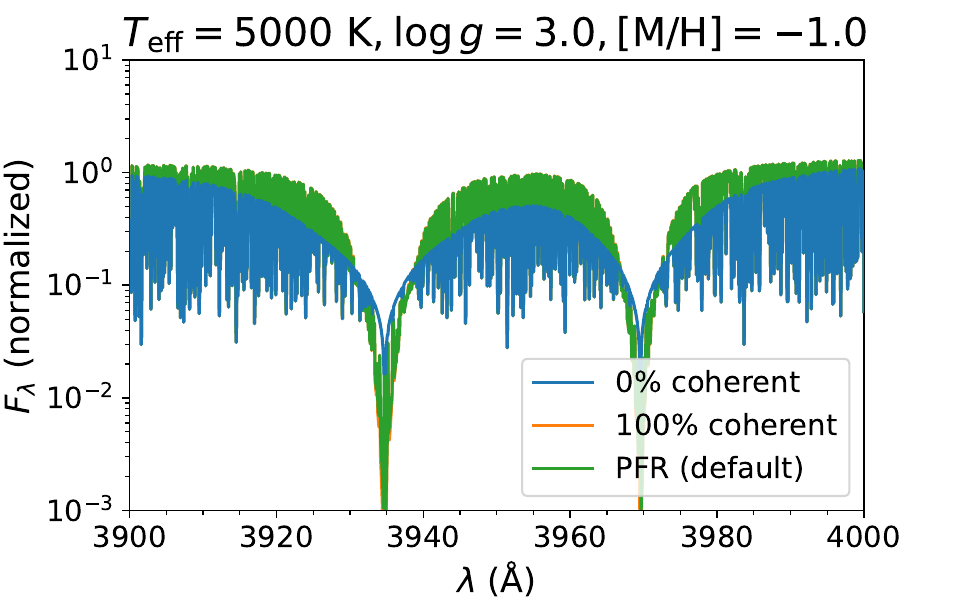}
    \includegraphics[width=0.99\linewidth]{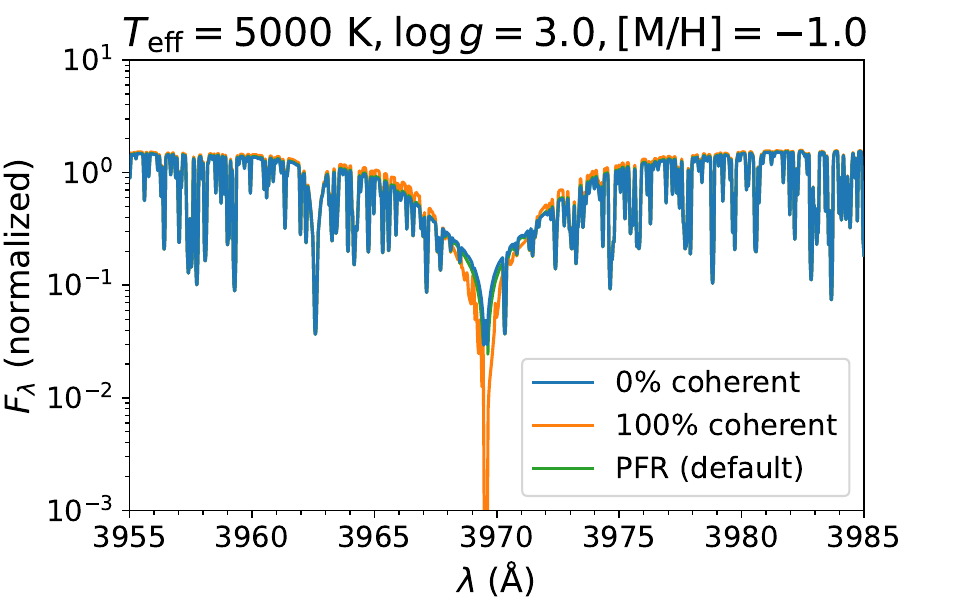}
    \caption{Effect of partial redistribution (PFR). Three cases are compared: treating the Ca HK line opacity as pure absorption (blue), pure scattering (orange), and a PFR treatment (green, the default in this work). Top: a low-gravity model. The PFR result closely aligns with the pure scattering case (the green line almost perfectly overlaps the orange), which predicts much narrower line profiles than the pure absorption case. Bottom: a high-gravity model (zoomed in on the Ca H line). The PFR result now more resembles the pure absorption case. }
    \label{fig:pfr}
\end{figure}
Stellar atmosphere models usually assume that the emission and absorption of photons in a bound-bound transition are independent. This will become inaccurate if, after an atom or ion absorbs a line photon and becomes radiatively excited, it has a significant probability of immediately re-emitting the photon by radiatively decaying to the original state. In this case, the frequencies of the absorbed and re-emitted photons become correlated, which requires the treatment known as partial frequency redistribution (PFR).

PFR in the form of the ``partial coherent scattering approximation'' has been implemented in \textsc{tlusty} for Lyman-$\alpha$ and the resonance lines of Mg I and II and in \textsc{synspec} for Lyman-$\alpha$ \citep[Appendix F]{Hubeny2017b}. We similarly implement PFR in \textsc{synspec} for the Ca HK lines. In brief, in the line wing (more than three Doppler widths away from the line center), we model a fraction of the line opacity as coherent scattering. We evaluate this coherence fraction at each spatial location, 
\begin{equation}
    \gamma_{\rm coh} = {\Gamma_R\over\Gamma_R + \Gamma_{\rm Stark} + \Gamma_{\rm vdW}}\,, \label{eq:coherence}
\end{equation}
where $\Gamma_R$, $\Gamma_{\rm Stark},$ and $\Gamma_{\rm vdW}$ are the natural, Stark, and van der Waals damping parameters of the line. Physically, $\Gamma_{\rm Stark}$ and $\Gamma_{\rm vdW}$ characterize the elastic collision rates of the excited Ca II ion with electrons and neutral hydrogen, which will lead to decoherence if they are much greater than the radiative decay rate, $\Gamma_R$. In the line core (less than three Doppler widths from the line center), we no longer introduce coherent scattering, which becomes negligible due to thermal frequency redistribution \citep{Jefferies1960}. We note that PFR is only relevant for resonance lines. For subordinate lines, such as the CaT, the excited state is far more likely to branch out to other channels of radiative decay (by emitting an H or K line photon in the case of Ca II) than to re-emit coherently. 

In Equation~(\ref{eq:coherence}), $\Gamma_{\rm Stark}$ and $\Gamma_{\rm vdW}$ are proportional to the electron and neutral hydrogen density, respectively, so we expect atmospheres with low $\log g$ (and therefore low photosphere density) to have higher $\gamma_{\rm coh}$. Figure~\ref{fig:pfr} shows the Ca HK profile with $\gamma$ given by Equation~(\ref{eq:coherence}) compared to setting $\gamma_{\rm coh}=0$ or $\gamma_{\rm coh}=1$ everywhere in the atmosphere. For $T_{\rm eff}=5000{\rm~K}, \log g=-2,\rm[M/H]=-1$, the PFR result is almost identical to the case with $100\%$ coherent scattering due to the low collision rates. In contrast, in the model with $\log g=3$, the much higher atmosphere density brings the PFR result closer to the case with $\gamma_{\rm coh}=0$. We note that the usual implicit assumption of $\gamma_{\rm coh}=0$ in stellar models will predict extremely broad absorption wings at low $\log g$, with a total EW of the Ca HK lines higher by $30\%$ than the PFR treatment.

As a caveat, we still assume that the Ca II level population satisfies LTE although PFR is essentially a non-LTE effect. Our treatment should be viewed as a first-order non-LTE correction to the LTE model; a fully self-consistent calculation will need to further account for statistical equilibrium. In addition, predicting the detailed line profile will likely require sophisticated frequency redistribution functions \citep[ch.~13,][]{Mihalas1978}, although coherent scattering is a good approximation in the far wing \citep{Jefferies1960}, which contributes most of the EW in low-$\log g$ models.

\section{Comparison with existing synthetic stellar spectra}
\label{app:validation}
\begin{figure}
    \centering
    \includegraphics[width=0.99\linewidth]{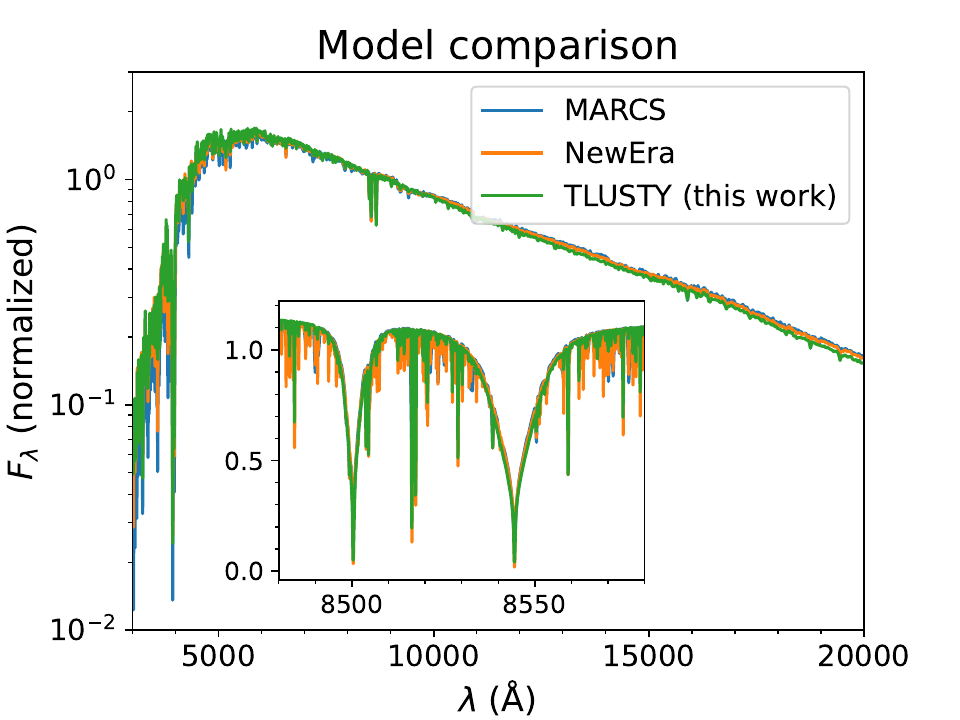}
    \caption{Comparison of synthetic spectrum in this work and two previously published stellar models. The parameters are $T_{\rm eff}=5000{\rm~K},\log g=0,{\rm[M/H]}=0$. The inset figure shows two of the CaT lines. Our result agrees with the published models in the overall SED and the CaT line profiles. }
    \label{fig:model_comparison}
\end{figure}

We test that our spectra calculated using \textsc{tlusty} agree with widely used theoretical spectral libraries in the stellar regime. Figure~\ref{fig:model_comparison} compares one example from our grid at $T_{\rm eff}=5000{\rm~K}, \log g=0, \rm[M/H]=0$ with spectra at the same parameter from two stellar grids in the literature, MARCS \citep{Gustafsson2008} and NewEra \citep{Hauschildt2025}. We find good agreement among the three spectra, with the continua different by $\lesssim10\%$ in the optical and $\lesssim5\%$ in the near-IR. Their absorption profiles of the CaT are also consistent, as shown in the inset. We note that MARCS and the current NewEra grid assume LTE, as in this work, but they use a spherical geometry designed for typical stellar radii. The atomic data are also different. These may explain the minor differences in the results.

\section{MCMC fitting for \textit{the Egg}}
\label{app:mcmc}

In this section, we describe our MCMC fitting methods and the results. As we have cautioned in the main text, the purpose of this exercise is to identify the global best fit and rule out potential degeneracies by scanning the parameter space. We refrain from interpreting the posteriors from the MCMC, which likely underestimate parameter uncertainties due to model systematics. 

We use uniform priors for all parameters as listed in Table~\ref{tab:MCMC}. We limit the parameter range of $\log g$ within our hydrostatic grid. We rebin the data and model spectra into 200 logarithmically equally spaced wavelength bins in $3200{\rm~\AA}<\lambda_{\rm rest}<22300{\rm~\AA}$. We also evaluate the synthetic photometry of the redshifted model spectrum in the WISE W1 band. The total logarithmic likelihood is given by
\begin{gather}
    \ln P = -{1\over2}\sum_{\rm bin}\left[{(L_{\lambda,\rm data}-L_{\lambda,\rm model})^2\over\sigma_{\rm \lambda,\rm data}^2+s^2L_{\lambda,\rm model}^2} + \right. \nonumber \\ 
    \left.\ln(\sigma_{\lambda,\rm data}^2+ s^2L_{\lambda,\rm model}^2)\right]\,, \label{eq:likelihood}
\end{gather}
where the sum goes over all the spectral wavelength bins (except for bins with strong emission lines in the data, which we mask off) and the W1 photometric band. The errors have two components: $\sigma_{\lambda,\rm data}$ comes from the spectral errors propagated through rebinning or from the WISE photometry, and $s$ is a free parameter representing the unknown relative systematic uncertainties in the model. We note that Equation~(\ref{eq:likelihood}) assumes that the errors in each bin are uncorrelated. This is the motivation of our rebinning the spectrum to low resolution: we found that the residuals would be strongly correlated if we directly used the original, high-resolution spectrum, which would result in severely overconfident posteriors. Therefore, we reduce the number of bins to 200 (or resolution $\sim100$), by which the residuals have a correlation length close to one bin (i.e., statistically uncorrelated; see also the residual in Figure~\ref{fig:egg_spectrum_model}). Rebinning also implies that we are mainly fitting the continuum shape rather than individual lines. However, rebinning does not strongly influence the best-fit parameter values. 

\begin{figure*}
    \centering
    \includegraphics[width=0.8\linewidth]{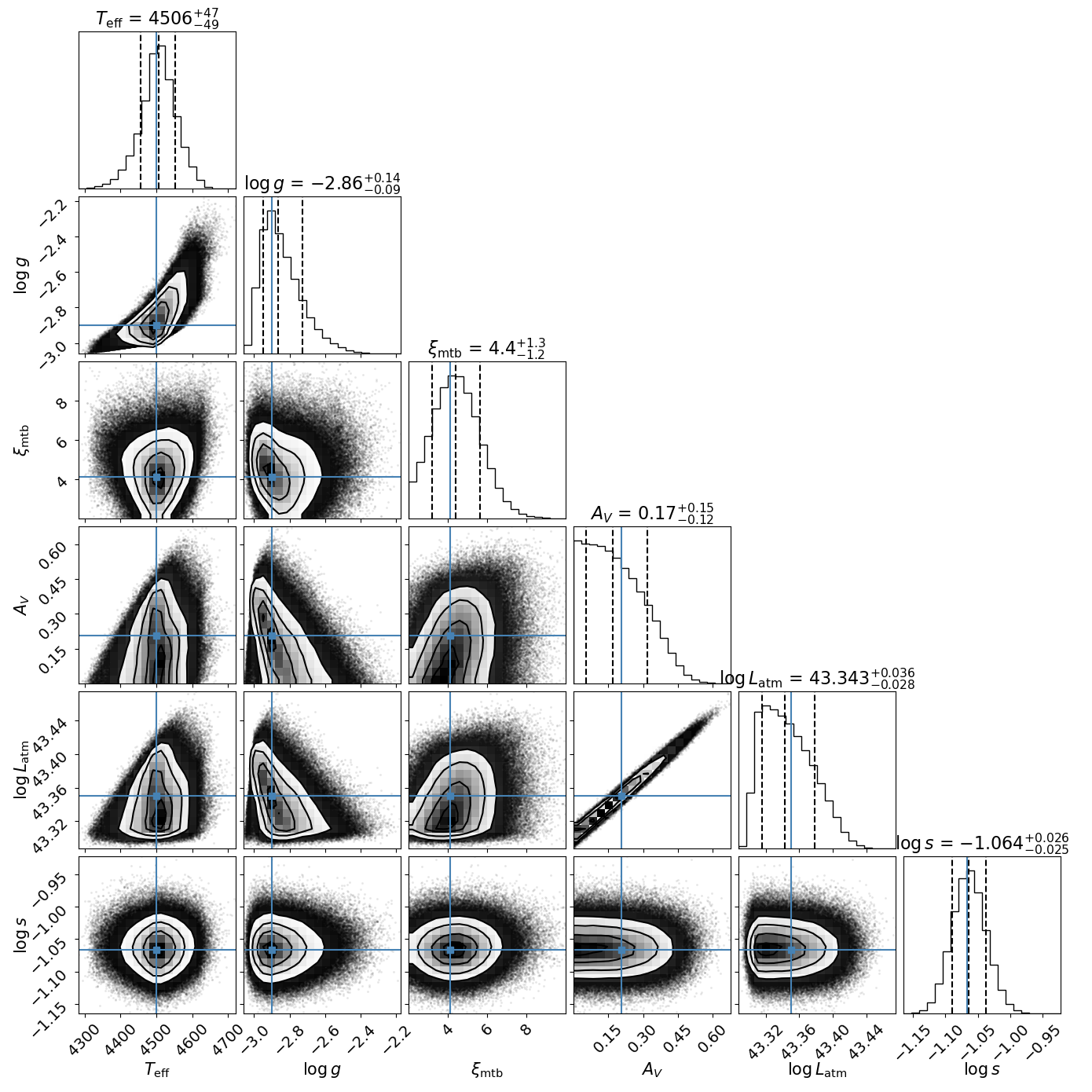}
    \caption{Corner plot of the MCMC posterior sample. Only parameters related to the atmosphere are shown.}
    \label{fig:MCMC_corner}
\end{figure*}

Figure~\ref{fig:MCMC_corner} shows the MCMC posterior of the atmosphere parameters and the uncertainty parameter. The best-fit is unique, and the posterior shows no sign of multi-modality. We further experimented artificially inflating the error at all wavelength bins by 10 to flatten the posterior, but still found no multi-modality. This supports the robustness of our inference of a low $\log g$. We also note that the best-fit $\log s$ implies that the average disagreement between the model-data is $\lesssim10\%$, which suggests reasonably good model performance. 

\section{Sensitivity tests for fitting to \textit{the Egg}}
\label{app:sensitivity}
\begin{table}
    \centering
    \begin{tabular}{lcc}
        \toprule
        MCMC run & $T_{\rm eff}$ posterior & $\log g$ posterior \vspace{1mm}\\
        \hline
        fiducial (Table~\ref{tab:MCMC}) & $4506^{+47}_{-49}$ & $-2.86^{+0.14}_{-0.09}$ \\ 
        $[\rm M/H]=0$\footnote{The grid is coarser than $\rm[M/H]=-1$ in the $T_{\rm eff}$ space and only includes $\xi_{\rm mtb}=2{\rm~km~s^{-1}}$ (Figure~\ref{fig:grid}).} & $4395^{+87}_{-91}$ & $-2.83^{+0.11}_{-0.10}$ \\
        $[\rm M/H]=-2$\textsuperscript{a} & $4497^{+10}_{-18}$ & $-2.78^{+0.05}_{-0.04}$ \\
        free $T_{\rm dust}$ & $4542^{+61}_{-67}$ & $-2.85^{+0.22}_{-0.11}$\\
        free $T_{\rm dust}$, no WISE & $4538^{+65}_{-64}$ & $-2.86^{+0.22}_{-0.10}$\\
        galaxy dust $A_{V\rm,gal}$ & $4517^{+35}_{-25}$ & $-2.94^{+0.07}_{-0.04}$ \\
        \bottomrule
    \end{tabular}
    \caption{Additional MCMC runs as sensitivity tests. Changing the assumed metallicity, allowing $T_{\rm dust}$ to vary, excluding the WISE photometry, or including dust extinction to the host galaxy does not change the main conclusion of a low $\log g$. }
    \label{tab:sensivitity}
\end{table}

We have made several simplifying parameter choices when modeling \egg\ in the main text. To show that our inference of low $\log g$ is robust, we report five additional MCMC runs where we relax some of those assumptions. The runs are summarized in Table~\ref{tab:sensivitity}. For each additional run, the first column describes the changed ingredient with all else held the same as fiducial (see also Table~\ref{tab:MCMC}). 

We first change the assumed atmospheric metallicity to $\rm[M/H]=0$ and $-2$, respectively. The posterior of $T_{\rm eff}$ and $\log g$ are consistent with that inferred with the fiducial $\rm[M/H]=-1$; the best-fit composite model spectra are similar across the optical-to-near-IR continuum. We caution that the grid is coarser in the $T_{\rm eff}$ space and only include $\xi_{\rm mtb}=2{\rm~km~s^{-1}}$ for these two metallicities compared to $\rm[M/H]=-1$. 

Then, we investigate the sensitivity to mid-IR dust emission by including $T_{\rm dust}$ as a free-varying parameter with a flat prior in $[300, 2000]{\rm~K}$. The fitting (4th row in Table~\ref{tab:sensivitity}) finds similar $T_{\rm eff}$ and $\log g$ posteriors to fiducial. The best-fit dust temperature is $8\times10^2{\rm~K}$, somewhat lower than the fiducial assumption, with a slightly higher best-fit dust luminosity of $\log(L_{\rm dust}/{\rm erg~s^{-1}})=42.9$. This mildly improves the match to the spectrum at $\lambda_{\rm rest}\sim2{\rm~\mu m}$ and the WISE W1 photometry; it still requires the atmosphere to produce the blue near-IR slope without a strong $\rm H^-$ kink, which is the main driver for the low $\log g$.

The fitting with free $T_{\rm dust}$ but without considering the WISE photometry shows posteriors similar to the free-$T_{\rm dust}$ run discussed above. This is because the many spectroscopic data points at $\lambda_{\rm rest}\sim2{\rm~\mu m}$ dominate over the single WISE photometric point in the likelihood. The best-fit dust component in this case is $T_{\rm dust}=9\times10^2{\rm~K}$ and $\log(L_{\rm dust}/{\rm erg~s^{-1}})=42.9$.  

Finally, we allow dust extinction to the host galaxy component, parameterized by $A_{V,\rm gal}$, with a flat prior between $0$ and $A_V$ (extinction to the atmosphere emission). The $T_{\rm eff}$ and $\log g$ posteriors are still similar to fiducial. This fit suggests moderate extinction for both the atmosphere and the host galaxy, with $A_{V,\rm gal}=0.37^{+0.08}_{-0.10}$ and $A_V=0.42^{+0.07}_{-0.08}$, and the residual is slightly improved in the UV than fiducial.

\bibliography{bib}{}
\bibliographystyle{aasjournalv7.1}
\end{CJK*}

\end{document}